\documentclass[preprint,12pt]{elsarticle}

\usepackage{amsmath,amsfonts,amssymb}
\usepackage{amsthm}

\usepackage{mathtools}
\usepackage[nice]{nicefrac} 
\usepackage{derivative}

\newtheorem{proposition}{Proposition}
\newtheorem{lemma}{Lemma}
\newtheorem{theorem}{Theorem}
\newtheorem{remark}{Remark}

\usepackage{subcaption}
\usepackage{graphicx}

\usepackage[nameinlink,capitalize]{cleveref}

\usepackage{circuitikz}

\usepackage{booktabs}

\newcommand{\Vpv}{V_\mathrm{pv}}
\newcommand{\Ipv}{I_\mathrm{pv}}
\newcommand{\Ppv}{P_\mathrm{pv}}

\newcommand{\Voc}{V_\mathrm{oc}}
\newcommand{\Isc}{I_\mathrm{sc}}
\newcommand{\Vmp}{V_\mathrm{mp}}
\newcommand{\Imp}{I_\mathrm{mp}}
\newcommand{\Pmp}{P_\mathrm{mp}}

\newcommand{\Iph}{I_\mathrm{ph}}
\newcommand{\Io}{I_\mathrm{o}}
\newcommand{\Rs}{R_\mathrm{s}}
\newcommand{\Gsh}{G_\mathrm{sh}}

\newcommand{\R}{\mathbb{R}}

\newcommand{\fpv}{f_\mathrm{pv}}
\newcommand{\tpv}{\theta_\mathrm{pv}}

\journal{Solar Energy}

\begin{document}

\begin{frontmatter}



\title{On the parameters domain of the single-diode model}


\author[inst1]{Carlos C\'ardenas-Bravo}
\author[inst3,inst4]{Denys Dutykh}
\author[inst2]{Sylvain Lespinats}

\affiliation[inst1]{organization={Univ. Grenoble Alpes, Univ. Savoie Mont Blanc, CNRS, LAMA},
            city={Chambéry},
            postcode={73000}, 
            country={France}}

\affiliation[inst2]{organization={Univ. Grenoble Alpes, CEA, Liten, INES},
            city={Le Bourget du Lac},
            postcode={73375}, 
            country={France}}

\affiliation[inst3]{organization={Mathematics Department},
            addressline={Khalifa University}, 
            city={Abu Dhabi},
            postcode={PO Box 127788}, 
            country={United Arab Emirates}}

\affiliation[inst4]{organization={Causal Dynamics Pty Lts},
            city={Perth},
            country={Australia}}

\begin{abstract}
The current-voltage (I-V) curves of solar photovoltaic (PV) systems have been widely used as a tool to determine their electrical operation. Usually, I-V curves are described employing three cardinal points: the short-circuit point $\left(0,I_\mathrm{sc}\right)$, the open-circuit point $\left(V_\mathrm{oc},0\right)$, and the maximum power point $\left(V_\mathrm{mp},I_\mathrm{mp} \right)$. In this context, to study the I-V curves, the single-diode model (SDM) is often used. In this manuscript, we present a practical approach for determining the parameters domain of the SDM given a set of cardinal points. An in-depth mathematical analysis showed that the SDM must satisfy the following inequalities $\nicefrac{V_\mathrm{oc}}{2} <V_\mathrm{mp}<V_\mathrm{oc}$ and $\nicefrac{I_\mathrm{sc}}{2}<I_\mathrm{mp}<I_\mathrm{sc}$. Then, it is possible to implicitly represent the SDM by a single parameter and, consequently, to determine its domain. For this purpose, a fast open-source computational algorithm for the single parameter domain of the SDM is presented. The algorithm has been computer-tested using $40\,000$ synthetically generated maximum power points which are processed in about one minute of the CPU time on a standard laptop computer. Moreover, as practical applications, the databases from the California Energy Commission and the NREL are analyzed using the presented algorithm.
\end{abstract}




\end{frontmatter}




\section{Introduction} 
\label{sec:intro}

\subsection{Fundamental concepts on the single-diode model} \label{subsec:fundamental}

\begin{figure}[!hbt]
\centering
\includegraphics[width=0.75\textwidth]{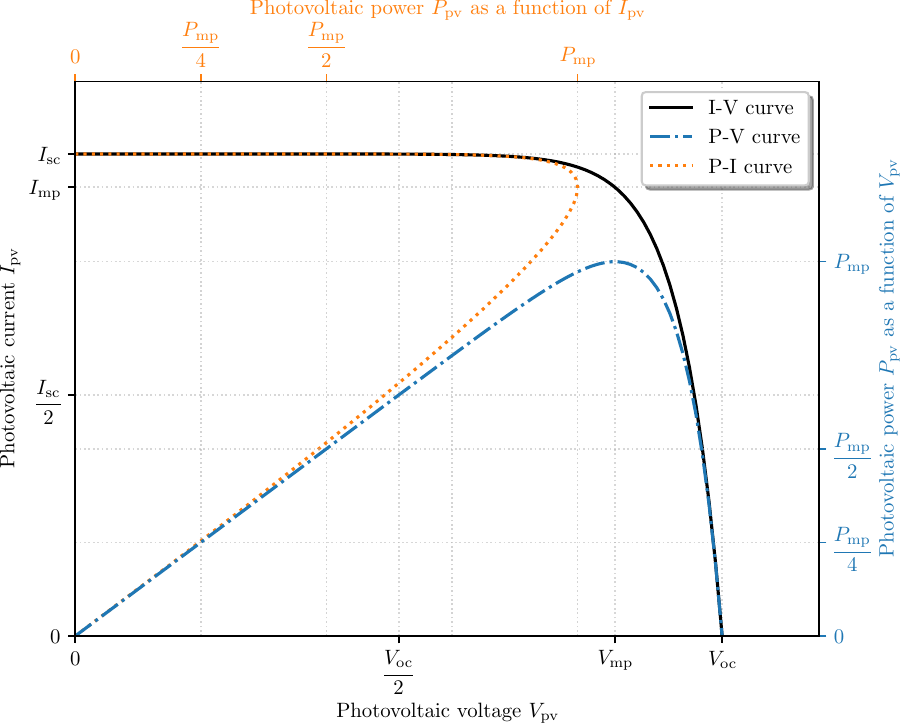}
    \caption{Typical I-V curve of a solar PV system. Here, each side of the figure represents a different axis: $\Ipv$ axis (left side), $\Vpv$ axis (bottom side), $\Ppv(\Vpv)$ axis (right side) and $\Ppv(\Ipv)$ axis (top side). For each curve there are different x- and y-axis: (i) the I-V curve is represented on the $\Ipv-\Vpv$ axes, (ii) the P-V curve is represented on the $\Vpv-\Ppv(\Vpv)$ axes and (iii) the P-I curve is represented on the $\Ipv-\Ppv(\Ipv)$ axes. } \label{fig:trial_ivcurve}
\end{figure}

The photovoltaic current $\Ipv$ and the photovoltaic voltage $\Vpv$ of a solar photovoltaic (PV) system are usually represented as a current-voltage (I-V) curve. 
From the I-V curve, three relevant operation points are identified, sometimes called remarkable points \citep{di_piazza_photovoltaic_2013} or cardinal points \citep{ndegwa_simplified_2020}: (i) short-circuit point (SCP), (ii) open-circuit point (OCP), and (iii) maximum power point (MPP). 
From these points are extracted the short-circuit current $\Isc$, the open-circuit voltage $\Voc$, the maximum power current $\Imp$, the maximum power voltage $\Vmp$, and the maximum power $\Pmp = \Vmp \Imp$.
\cref{fig:trial_ivcurve} shows with a solid black line the typical I-V curve. 
From the I-V curve, it is possible to extract the Power-Voltage (P-V) curve, presented in a blue dash-dotted line on \cref{fig:trial_ivcurve}. 
The y-axis of the P-V curve is represented by the right side of the figure. 
Complementary to the P-V curve, it is also possible to represent the behavior of the power as a function of the current (P-I curve). 
The P-I curve is presented with a dotted orange line on \cref{fig:trial_ivcurve}. 
The x- and y-axis of the P-I curve are represented by the left and top sides of the figure, respectively.

To better understand the I-V curves on solar PV systems, mathematical models have been developed. 
An early approximation is presented in the literature \citep{pfann_radioactive_1954,prince_silicon_1955} consisting of two components: (i) a current source of magnitude $\Iph$ proportional to the light intensity and (ii) a reverse current flowing internally through the solar cell junction \citep{rauschenbach_solar_1980}, sometimes denominated as the dark current \citep{nelson_physics_2004}. 
In turn, the dark current describes the diffusion of minority carriers from the neutral regions to the depletion region (dark saturation current $\Io$) \citep{dorf_electrical_2006} and the internal recombination process of the solar cell (ideality factor of the diode $n$) \citep{caprioglio_origin_2020}. 
For this work, the model comprising $\Iph$, $\Io$, and $n$ is denominated the Null Resistive Loss Model (NRLM).

Two additional resistive elements can be introduced to the early model described \citep{rauschenbach_solar_1980}. The first one is called series resistance $\Rs$ and takes into account the effect of the bulk resistance of the junction, the contact resistance between the junction and the electrodes, and the resistance of the electrodes themselves \citep{smets_solar_2016}. The other resistance is called shunt resistance $R_\mathrm{sh}$ and incorporates the effect of all non-linear currents flowing through the ohmic path across the n-p junction \citep{breitenstein_shunt_2004}. The model comprising $\Rs$ is denominated as Null Shunt Conductance Model (NSHM). Similarly, the model comprising $R_\mathrm{sh}$ is denominated as Null Series Resistance Model (NSRM).

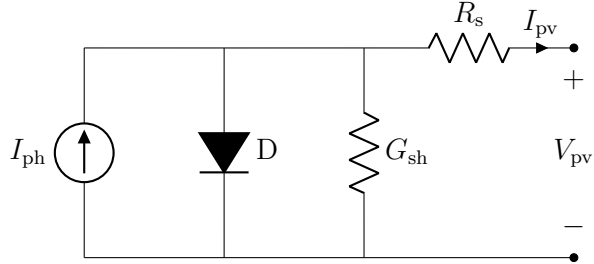
\begin{figure}[!hbt]
    \centering
    \begin{tikzpicture}[american,scale=0.93,every node/.style={transform shape}]
        \draw (0,0) to[current source,l=$\Iph$] (0,3);
        \draw (2,3) to[full diode,l=D] (2,0);
        \draw (4,3) to[R,l=$\Gsh$] (4,0);
        \draw (4,3) to[R,-*,l=$\Rs$,i=$\Ipv$] (7,3) to[open,v=$\Vpv$] (7,0) to[short, *-] (4,0);
        \draw (0,0) to[short] (4,0);
        \draw (0,3) to[short] (4,3);
    \end{tikzpicture}
    \caption{Photovoltaic SDM for a solar cell. The parameters of this model are distributed into four electrical elements: the current source ($\Iph$ parameter), the diode $D$ ($\Io$ and $A$ parameters), and two resistances ($\Rs$ and $\Gsh$ parameters).} \label{fig:sdm_circuit}
\end{figure}

In this context, the single-diode model is usually presented as shown by \cref{fig:sdm_circuit}. The main equation is
\begin{equation}
    \label{eq:sdm_a}
    \Iph = \Io \left( \exp{\left( \frac{ \Vpv + \Rs \Ipv}{ A} \right)} -1 \right) + \Gsh  (\Vpv + \Rs  \Ipv) + \Ipv,
\end{equation}
where $A$ represents the equivalent factor of the diode and is related to the ideality factor of the diode as
\begin{equation}
    \label{eq:ideality}
    n \coloneqq  A \frac{q}{k_\mathrm{B}  T_\mathrm{s}}.
\end{equation}
In this equation $q$ is the charge of the electron, $k_\mathrm{B}$ is the Boltzmann constant and $T_\mathrm{s}$ is the temperature of the PV system. Furthermore, if $N_\mathrm{p}$, $N_\mathrm{s}$, are known, the parameters of the solar cell can be calculated as indicated in \cref{app_sdm_solarcell}. For this, the following assumptions are required \citep{femia_pv_2013}: (i) uniform irradiance, (ii) uniform cell temperature, and (iii) same state of health over the sections of the device. 

There are other high complex models such the case of the double-diode model (DDM).
This model, as it name indicates, counts on two diodes instead of one as in the SDM.
The incorporation of the second diode allows us to have a better description of the recombination in the solar cell junction.
Advantages of this model are a better description for low irradiances scenarios.
However, for this work, the SDM will be used as an example.
Therefore, the next section explain the currents solutions for the SDM.

\subsection{Solutions for the SDM}

With the purpose of identifying the parameters of the equation, most of the authors agree that there are mainly two kinds of solution: (i) analytic solution, and (ii) numerical solution. Analytical algorithms explicitly express the solution of the model \citep{merchaoui_particle_2018,de_blas_selecting_2002}, and are also denominated as noniterative, direct, or explicit algorithms \citep{batzelis_non-iterative_2019}. The advantage of the direct algorithm lies in its simplicity and fast calculation. However, in general, several approximations have to be adopted. In this context, Phang \emph{et al.} and Chang \emph{et al.} reported a series of works describing the first non-iterative method to determine the SDM \citep{phang_accurate_1984,chan_comparative_1986}. Its solution consisted of using the slope of the I-V curve at the short-circuit point as a way to approximate $R_\mathrm{sh}$. In addition, the slope of the I-V curve evaluated at the open-circuit voltage is used, as indicated by \citep{kennerud_analysis_1969}. As a result, five explicit expressions can be found to express the parameters of the SDM. This solution had to be described because the nonlinear equation system was very dependent on the initial point. Later analytical solutions exposed in the state-of-the-art uses the Lambert W function, first applied in the context of photovoltaics by Jain and Kapoor \citep{jain_exact_2004}. Nassar-eddine \emph{et al.} presented an extraction method based on the Lambert W function \citep{nassar-eddine_parameter_2016}. This method is based on using the temperature coefficient of the open-circuit voltage $\beta_\mathrm{oc}$. Then, an explicit expression of the open-circuit voltage as a function of the temperature must be determined. However, to express $\Voc$ as a function of the temperature, the thermal behavior of the different parameters must be determined. This fact supposes a challenge since extra equations are incorporated from the thermal theory (including undesired errors on the calculated parameters). 

The second kind of solution is the numerical solution. This solution can be divided into deterministic or stochastic \citep{chin_coyote_2019}. Numerical deterministic methods usually determine the solution of an equation system, \emph{e.g.}, the Levenberg-Marquardt algorithm. Easwarakhanthan \emph{et al.} proposed an early method to calculate the five parameters of the SDM equation based on the Newton model modified with the Levenberg parameter \citep{easwarakhanthan_nonlinear_1986}. 
In his work, to initialize its algorithm, the parameters $\Iph$, $\Io$ and $\Gsh$ are calculated as a linear regression as a function of the parameters $A$ and $\Rs$, previously indicated by Nguyen \emph{et al.} \citep{nguyen_computer-aided-characterization_1982}. 
In this line, several works have demonstrated that the parameters of the PV model depend on the temperature \citep{varshni_temperature_1967,ahmad_effects_1992} and the irradiance \citep{townsend_method_1989}. 
An important work that exposes a solution of the SDM incorporating the translation of the parameters is that indicated by De Soto \emph{et al.} \citep{de_soto_improvement_2006}. 
De Soto \emph{et al.} proposed determining the five parameters by evaluating the cardinal points of the I-V curves and the temperature coefficient of the open-circuit voltage $\beta_{\mathrm{oc}}$. The limitation of this methodology is that it is mandatory to know the value of $\beta_{\mathrm{oc}}$.

Numerical stochastic methods (heuristic/meta-heuristic in this context) have been exploited in the past few years as mechanisms to identify the parameters of the SDM. Meta-heuristic algorithms can be classified into evolutionary algorithms, physics-based algorithms, swarm-based algorithms, and human-based algorithms \citep{oliva_review_2019}. These algorithms require minimizing an objective function, for example, the mean square error of the PV current \citep{younis_comprehensive_2022}. Usually, these algorithms,\textit{ e.g.}, differential evolution, require the knowledge of the domain for the parameters under optimization. For this purpose, the SDM can be expressed as a function of three variables using the approach described by Laudani \textit{et al.} as explained below.

In a brief note, Laudani \emph{et al.} presented a way to express the SDM as a function of two variables: (i) the ideality factor of the diode and (ii) the series resistance \citep{laudani_reduced-form_2013}. In the same line, Laudani described the boundaries of both variables in order to guarantee the convergence of the resolution algorithm to the right solution. The definition of parameter boundaries is a major challenge to find adequate solutions for the SDM. The lower boundaries for all five variables must be strictly positive to ensure a model that makes physical sense. However, with respect to the upper boundaries, these must be analyzed individually due to the different nature of the parameters. The upper boundary of the ideality factor is generally accepted as 2 \citep{de_soto_improvement_2006,oliva_review_2019,ishaque_improved_2011,niu_biogeography-based_2014}. On the other hand, the upper boundary for the series resistance, Laudani proposed an analytic expression for practical I-V curves considering that the maximum power voltage is about 80\% of the open-circuit voltage. This supposition simplifies some expressions in the mathematical approach, allowing one to explicitly find the maximum boundaries for the series resistance. After that, Laudani proposed a fast calculation algorithm to extract the SDM \citep{laudani_very_2014} using his previous work as the basis. To demonstrate the effectiveness of the algorithm, he used the California Energy Commission (CEC) database. On the same line, two related works have been published that provide additional details on this new methodology and how to use it on real data \citep{laudani_high_2014,laudani_identification_2014}. After some years, in 2019 Coco \emph{et al.} \citep{coco_sensitivity_2019} performed a sensitivity analysis of the reduced form proposed by Laudani. Coco \emph{et al.} found that a good value $n$ is located near its upper limit. Toledo \emph{et al.} \citep{toledo_geometric_2014,toledo_-depth_2021,toledo_quick_2022} provided a depth analysis to determine the SDM.

In addition to the analytic and numerical solutions, Machine Learning (ML) approaches have been used to estimate relevant parameters describing the I-V response \citep{jobayer_systematic_2023}. Examples of these algorithms are presented by Wang \textit{et at.} \citep{wang_accurate_2021} and Zhang \textit{et al.} \citep{zhang_performance_2020}. These algorithms determine directly the optimal models used to describe the I-V curve, however, no further information can be extracted regarding the boundaries of the parameters. 

\subsection{Contributions}

In this work we describe the single-diode model by two implicit functions in a thoroughly chosen coordinate space using the three cardinal points: $(0,\Isc)$, $(\Vmp, \Imp)$, and $(\Voc,0)$. In the first part, we demonstrate that the SDM must satisfy $\nicefrac{\Voc}{2}<\Vmp<\Voc$ and $\nicefrac{\Isc}{2}<\Imp<\Isc$ (\emph{cf.} \cref{theorem:mppSDM}). In the second part, we show that there exist a set of SDM satisfying these cardinal points defined implicitly by one single parameter (\emph{cf.} \cref{theorem_one-dimensional}). Therefore, it is possible to determine the SDM parameter domain, given any set of cardinal points (extracted from field measurements or provided by the manufacturer). Then, by using \cref{theorem:mppSDM,theorem_one-dimensional} a fast calculation algorithm to determine the parameter domain of the implicit SDM is proposed in this study (the source code can be found on \cref{sec:conclusion}). This algorithm has been tested on synthetic and real data sets demonstrating its applicability. 
The implicit definition allows a fast determination of the parameters of the closest SDM of a measured I-V curve.

\subsection{Outline}

In \cref{sec:general_considerations}, the main mathematical definitions used are introduced. \cref{sec:maximum_power_point_SDM} presents a detailed description of the procedure to express $\Iph$ and $\Io$ as functions of $A$, $\Rs$, and $\Gsh$. 
From this, the boundaries of the maximum power point of the SDM are found (\emph{cf.} \cref{theorem:mppSDM}). \cref{sec:feasible_domain} explains how to express $\Iph$, $\Io$ and $\Gsh$ as functions of $(0,\Isc)$, $(\Voc,0)$, and a point $(V_\mathrm{x},I_\mathrm{x})$ selected according to \cref{lemma:concavity}. 
Then, given a set of three cardinal points, there is a series resistance function defining implicitly the SDM (\emph{cf.} \cref{theorem_one-dimensional}).
This methodology, applied to the SDM, is generalized in \cref{sec:methodology_description}.
\cref{sec:comp_alg} indicates a simple and fast calculation method for the limits of the one-parameter SDM. Using the calculation algorithm, in \cref{sec:results_scaled} the maximum limit of the equivalent factor of the diode $A_\mathrm{max}$ is calculated for an artificial data set. Then, in \cref{sec:results_cec} and \cref{sec:results_nrel}, real databases are analyzed to demonstrate the practical applicability of the presented approach.

\section{Main definitions of the SDM} 
\label{sec:general_considerations}
This section provides the parametric representation of the SDM (\emph{cf.} \cref{ssec:parametric_representation}). Then, from this definition, relevant mathematical properties of the SDM are provided.
\subsection{Parametric representation} \label{ssec:parametric_representation}
Let express \cref{eq:sdm_a} as a function $F_\mathrm{SD}(X;\Theta)=0$, indicated in \cref{eq:Fsdm}. The inputs of $F_\mathrm{SD}$ are represented by $X \coloneqq \{ V_\mathrm{pv},I_\mathrm{pv} \}$ such that $X\in\mathbb{R}^2$. The set of parameters of $F_\mathrm{SD}$ are represented by $\Theta \coloneqq \{ I_\mathrm{ph},I_\mathrm{o},A,R_\mathrm{s},G_\mathrm{sh} \}$ such that $\Theta \in \mathbb{R}_+^5$. In this paper, the term dimensionality refers exclusively to the number of model parameters. Then, \cref{def:SDM} presents the formal definition of the SDM.
\begin{equation}
    \label{eq:Fsdm}
    F_\mathrm{SD}(X ; \Theta) \coloneqq I_{\mathrm{o}} \exp{ \left( \frac{ V_{\mathrm{pv}}+R_{\mathrm{s}}  I_{\mathrm{pv}}}{A}  \right)} + G_{\mathrm{sh}}  ( V_{\mathrm{pv}}+R_{\mathrm{s}}  I_{\mathrm{pv}} ) + I_{\mathrm{pv}} - (I_\mathrm{ph}+I_\mathrm{o}) 
\end{equation}
\begin{equation}
    \label{def:SDM}
    \text{SDM} \coloneqq \{ (X,\Theta) \in \mathbb{R}^2 \times \mathbb{R}_+^5 \mid  F_\mathrm{SD}(X ;\Theta) = 0  \}
\end{equation}

To better understand the SDM, the NRLM, NSHM, and NSRM introduced in \cref{subsec:fundamental} are used as a reference in the analysis (\cref{def:NRLM,def:NSHM,def:NSRM}). Under these definitions, \cref{prop:NRLM_mpp_boundaries} is presented.
\begin{equation}
    \label{def:NRLM}
    \text{NRLM} \coloneqq \{ (X,\Theta) \in \mathbb{R}^2 \times \mathbb{R}_+^3 \times \{ 0 \}^2 \mid F_\mathrm{SD}(X ;\Theta) = 0  \}
\end{equation}
\begin{equation}
    \label{def:NSHM}
    \text{NSHM} \coloneqq \{ (X,\Theta) \in \mathbb{R}^2 \times \mathbb{R}_+^3 \times \{ 0 \} \times \mathbb{R}_+ \mid  F_\mathrm{SD}(X ;\Theta) = 0  \}
\end{equation}
\begin{equation}
    \label{def:NSRM}
    \text{NSRM} \coloneqq \{ (X,\Theta) \in \mathbb{R}^2 \times \mathbb{R}_+^4 \times \{ 0 \} \mid F_\mathrm{SD}(X ;\Theta) = 0  \}
\end{equation}
\begin{proposition} \label{prop:NRLM_mpp_boundaries}
    For a given set of parameters $\{ I_\mathrm{ph},I_\mathrm{o},A\}$, 
    the photovoltaic power determined by the SDM, the NSHM, and the NSRM is lower than the NRLM.
\end{proposition}
\begin{proof}
    Let $P_\mathrm{RL}$ be the output power of the NRLM such that $P_\mathrm{RL}>0$. The incorporation of any extra external resistance on the NRLM will decrease the $P_\mathrm{RL}$. Since the SDM, the NSHM, and the NSRM can be reduced as the NRLM and an external resistive system, then the MPP must be maximum for the NRLM.    
\end{proof}
\subsection{Concavity of the model}
The I-V curve from the SDM it is said to be concave down (decreasing) on $\Vpv$. This property was previously demonstrated by Toledo et al. \cite{toledo_geometric_2014} (see details in \cref{app:concavity}). A direct consequence of this property is provided in \cref{lemma:concavity}. In addition, from \cref{lemma:concavity} it is extracted \cref{prop:minimum_power}, which imposes conditions on the cardinal points of the SDM.
\begin{proposition} \label{lemma:concavity}
The SDM is such that $\Vpv  I_\mathrm{sc} + V_\mathrm{oc}  \Ipv \geqslant V_\mathrm{oc}  I_\mathrm{sc} \, \forall \, 0 \leqslant \Vpv \leqslant V_\mathrm{oc}$.
\end{proposition}
\begin{proof}
\par Since $\odif[switch-*=true]{\Vpv}{\Ipv}<0$ and $\odif[switch-*=true,order=2]{\Vpv}{\Ipv}<0$, then  $\Vpv  I_\mathrm{sc} + V_\mathrm{oc}  \Ipv \geqslant V_\mathrm{oc}  I_\mathrm{sc} \, \forall \, 0 \leqslant \Vpv \leqslant V_\mathrm{oc}$.
\end{proof}
\begin{lemma} \label{prop:minimum_power}
The cardinal points of the SDM are such that $\nicefrac{V_\mathrm{oc}}{2}<V_\mathrm{mp}<V_\mathrm{oc}$ and $\nicefrac{I_\mathrm{sc}}{2}<I_\mathrm{mp}<I_\mathrm{sc}$.
\end{lemma}
\begin{proof}
\par Let $X$ be the solution of $F_\mathrm{SD}$ with parameters $\Theta$, such that $\Vpv,\Ipv>0$. Now, let's consider a random point $X^\mathrm{rng}=(\Vpv^\mathrm{rng},\Ipv^\mathrm{rng})$ on $X$. Then, it is possible to approximate the original I-V curve by: (i) one segment connecting $(0,I_\mathrm{sc})$ and $X^\mathrm{rng}$ and (ii) another segment connecting $X^\mathrm{rng}$  and $(V_\mathrm{oc},0)$. The maximum power point of the approximated I-V curve $(V_\mathrm{mp}^\mathrm{rng},I_\mathrm{mp}^\mathrm{rng})$ belongs to the interval shown by \cref{interval:mp} (see details in \cref{appendix:cp-limits}). Then, it is accomplished $V_\mathrm{mp}>\nicefrac{V_\mathrm{oc}}{2}$ and $I_\mathrm{mp}>\nicefrac{I_\mathrm{sc}}{2}$. Moreover, since the SDM is concave down (decreasing) on $\Vpv$, it follows $V_\mathrm{mp}<V_\mathrm{oc}$ and $I_\mathrm{mp}<I_\mathrm{sc}$.
\begin{equation}
\label{interval:mp}
V_\mathrm{mp}^{\mathrm{rng}} \in \, \left]\frac{V_\mathrm{oc}}{2},V_\mathrm{mp} \right[  \quad \text{and} \quad I_\mathrm{mp}^{\mathrm{rng}} \in \, \left] \frac{I_\mathrm{sc}}{2},I_\mathrm{mp} \right[ 
\end{equation}
\end{proof}
\subsection{Maximum power point condition}
\begin{proposition} \label{lemma:photovoltaic_mpp}
The maximum power point of the SDM is such that $ \odif[switch-*=true]{\Vpv}{\Ipv} \big|_{\mathrm{mp}}\coloneqq-\nicefrac{I_\mathrm{mp}}{V_\mathrm{mp}}$ and $\odif[switch-*=true]{\Ipv}{\Vpv} \big|_{\mathrm{mp}} \coloneqq -\nicefrac{V_\mathrm{mp}}{I_\mathrm{mp}}$. 
\end{proposition}
\begin{proof}
The electrical power $P_\mathrm{pv} \coloneqq \Vpv  \Ipv$ follows $\odif[switch-*=true]{\Vpv}{P_\mathrm{pv}} \coloneqq \Ipv + \Vpv  \odif[switch-*=true]{\Vpv}{\Ipv}$ and $\odif[switch-*=true]{\Ipv}{P_\mathrm{pv}} \coloneqq \Vpv + \Ipv  \odif[switch-*=true]{\Ipv}{\Vpv}$. Moreover, $P_\mathrm{mp}$ evaluated at the MPP accomplishes $\odif[switch-*=true]{\Vpv}{P_\mathrm{pv}} \big|_{\mathrm{mp}}=\odif[switch-*=true]{\Ipv}{P_\mathrm{pv}} \big|_{\mathrm{mp}} = 0 $. Therefore, $\odif[switch-*=true]{\Vpv}{\Ipv} \big|_{\mathrm{mp}} \coloneqq-\nicefrac{I_\mathrm{mp}}{V_\mathrm{mp}}$ and $\odif[switch-*=true]{\Ipv}{\Vpv} \big|_{\mathrm{mp}} \coloneqq -\nicefrac{V_\mathrm{mp}}{I_\mathrm{mp}}$. 
\end{proof}

\section{Maximum power point of the SDM} \label{sec:maximum_power_point_SDM}
This section presents a way to represent the single diode model as a function of the set of parameters $\Theta_3 \coloneqq \{ A, \,R_\mathrm{s}, \,G_\mathrm{sh} \}$ given $(0,\,I_\mathrm{sc})$ and $(V_\mathrm{oc},\,0)$. Then, different $\Theta_3$ will produce different MPPs in such a way that a region in the I-V plane is described  (\emph{cf.} \cref{theorem:mppSDM}). 
\subsection{Three-parameters SDM representation} \label{sec:3p_sdm_representation}

Consider the points $(0,\,I_\mathrm{sc})$ and $(V_\mathrm{oc},\,0)$ given. Then, $I_\mathrm{ph}$ and $I_\mathrm{o}$ can be expressed as functions of $\Theta_3$. Replacing $I_\mathrm{ph}(\Theta_3)$ and $I_\mathrm{o}(\Theta_3)$ on \cref{def:SDM}, the three-parameter SDM (SDM-3) is formally defined as indicated by \cref{def:SDM-3}. Here, $F_\mathrm{SD3}(X;\,\Theta_3)$ is indicated by \cref{eq:sdmmodel003} (see details in \cref{appendix:3D_SDM}).
\begin{equation}
\text{SDM-3} \coloneqq \{ (X,\Theta_3) \in \mathbb{R}^2 \times \mathbb{R}_+^3  \mid F_\mathrm{SD3}(X;\Theta_3) = 0 ,\, \{ I_\mathrm{ph}(\Theta_3) ,\, I_\mathrm{o}(\Theta_3) \} >0 \} \label{def:SDM-3}
\end{equation}
%
%
Since $I_\mathrm{ph}(\Theta_3)$ and $I_\mathrm{o}(\Theta_3)$ must be positives, novel restrictions on $R_\mathrm{s}$ and $G_\mathrm{sh}$ are incorporated to the original model. Then, from the definition of the SDM-3, lower dimensional representations for the NRLM, NSHM, and NSRM are found, indicated by \cref{def:NRLM-1,def:NSHM-2,def:NSRM-2}. Here, $T_\mathrm{s}$ represents the limits of $R_\mathrm{s}$ whereas $T_\mathrm{sh}$ represents the limits of $G_\mathrm{sh}$, indicated by \cref{eq:bounds} (see details in \cref{appendix:3D_SDM}). It is important to note that the maximum limit of $G_\mathrm{sh}$ is a function of $R_\mathrm{s}$, therefore its value must be selected accordingly. Using these lower dimensional models, the region of the maximum power point of the SDM is determined.
\begin{multline}
\text{NRLM-1} \coloneqq \{ (X,\Theta_3) \in \mathbb{R}^2 \times \mathbb{R}_+ \times \{ 0 \}^2 \mid \\ F_\mathrm{SD3}(X,\Theta_3) = 0 , \, \{ I_\mathrm{ph}(\Theta_3) ,\, I_\mathrm{o}(\Theta_3) \} >0 \} \label{def:NRLM-1}
\end{multline}
\begin{multline}
\text{NSHM-2} \coloneqq \{ (X,\Theta_3) \in \mathbb{R}^2 \times \mathbb{R}_+^2 \times \{ 0 \} \mid \\ F_\mathrm{SD3}(X;\Theta_3) = 0 , \, \{ I_\mathrm{ph}(\Theta_3) ,\, I_\mathrm{o}(\Theta_3) \} >0   \} \label{def:NSHM-2} 
\end{multline}
\begin{multline}
\text{NSRM-2} \coloneqq \{ (X,\Theta_3) \in \mathbb{R}^2 \times \mathbb{R}_+ \times \{0\} \times \mathbb{R}_+ \mid \\ F_\mathrm{SD3}(X;\Theta_3) = 0 , \, \{ I_\mathrm{ph}(\Theta_3) ,\, I_\mathrm{o}(\Theta_3) \} >0 \} \label{def:NSRM-2} 
\end{multline}
\begin{align}
T_\mathrm{s} \coloneqq \biggr]0, \, \frac{V_\mathrm{oc}}{I_\mathrm{sc}} \biggl[ \quad \text{and} \quad T_\mathrm{sh} \coloneqq \biggr] 0, \, \frac{I_\mathrm{sc}}{V_\mathrm{oc}-R_\mathrm{s}  I_\mathrm{sc}} \biggl[ \label{eq:bounds}
\end{align}
\subsection{Maximum power point region} \label{sec:maximum_power_point_region}
\begin{figure}[!hbt]
\centering
\includegraphics[width=0.75\textwidth]{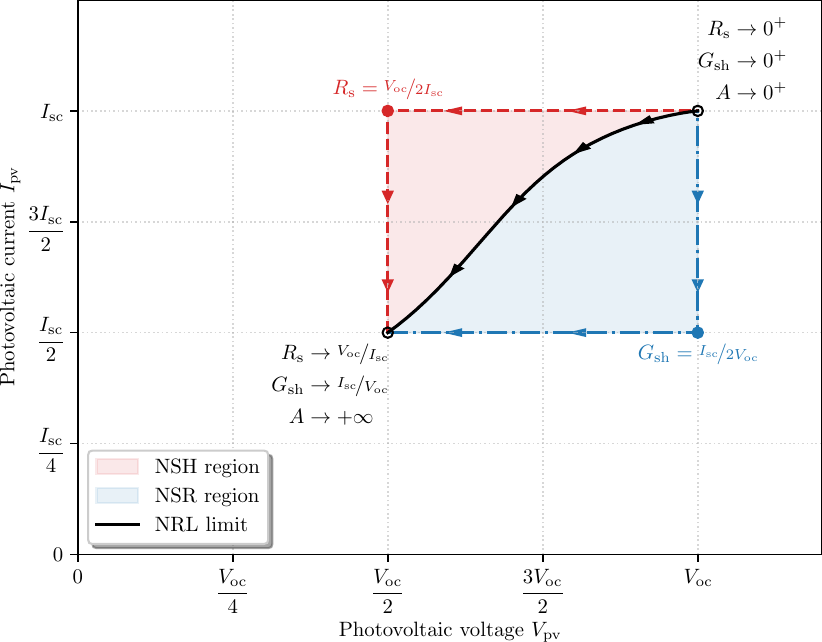}
\caption{Maximum power point region for the SDM. It is appreciated: (i) the NSH region (red area), (ii) the NSR region (blue area), and (iii) the NRL limit (black line). On the points $\left( V_\mathrm{oc},\,I_\mathrm{sc} \right)$ and $\left( \nicefrac{V_\mathrm{oc}}{2},\, \nicefrac{I_\mathrm{sc}}{2} \right)$ are indicated the values of the parameters for the NRLM, the NSHM and the NSRM. The dashed red line indicates the MPP of the NSHM for $A\rightarrow0^+$ and $R_\mathrm{s} \in \left]0,\, \nicefrac{V_\mathrm{oc}}{I_\mathrm{sc}}\right[$. Similarly, the dot-dashed blue line indicated the evolution of the MPP of the NSRM for $A\rightarrow 0^+$ and $G_\mathrm{sh} \in \left] 0,\, \nicefrac{I_\mathrm{sc}}{V_\mathrm{oc}} \right[$. For all cases, the arrows indicate the increasing direction of $R_\mathrm{s}$ or $G_\mathrm{sh}$. } \label{fig:theorem001}
\end{figure}

The MPP of the SDM belongs to a region defined by the NSHM-2 and NSRM-2 (\emph{cf.} \cref{theorem:mppSDM}). In this sense, this region can be divided into two sub-regions: (i) the NSH region (\emph{cf.} \cref{lemma:mppNSHM}) and (ii) the NSR region (\emph{cf.} \cref{lemma:mppNSRM}). In turn, the NSH region and the NSR region share a common boundary, called NRL limit (\emph{cf.} \cref{lemma:NRLlimit}). \cref{fig:theorem001} illustrates the NSH region, the NSR region, and the NRL limit. Here, the MPP of any SDM given $\Theta \in \mathbb{R}_+$ must be inside the region described by the union of the NSH region (red area) and the NSR region (blue area). 
\begin{proposition} \label{prop:mpp004}
The maximum power of the SDM decreases with the equivalent factor of the diode $A$.
\end{proposition}
\begin{proof}
Let be a SDM with $\Theta>0$ and $I_\mathrm{ph}$, $I_\mathrm{o}$, $R_\mathrm{s}$, and $G_\mathrm{sh}$ fixed. According to \cref{eq:1stderivativeIpvSC3D_dA}, $\odif[switch-*=true]{V_\mathrm{pv}}{I_\mathrm{pv}}$ is a monotonically decreasing function of $A$ at the short-circuit point. Then, since the SDM is concave down (decreasing), if $A_\mathrm{1}<A_\mathrm{2}$ then it is accomplished $I_\mathrm{pv}(A_\mathrm{1})>I_\mathrm{pv}(A_\mathrm{2})$ for $V_\mathrm{pv} \in \left]0, V_\mathrm{oc} \right[ $. Consequently, it is satisfied $P_\mathrm{mp}(A_\mathrm{1})>P_\mathrm{mp}(A_\mathrm{2})$.
\begin{equation}
    \partial_A \Biggl[\odif[switch-*=true]{V_\mathrm{pv}}{I_\mathrm{pv}} \bigg\rvert_{(V_\mathrm{pv},\,I_\mathrm{pv})=(0,\,I_\mathrm{sc})} \Biggr] \coloneqq \cfrac{ I_\mathrm{o}  \left( 1 + \cfrac{R_\mathrm{s}  I_\mathrm{sc}}{A} \right)  \exp{ \left( \cfrac{ R_\mathrm{s}  I_\mathrm{sc} }{A} \right) } }{ \left( I_\mathrm{o}  R_\mathrm{s}  \exp{ \left( \cfrac{ R_\mathrm{s}  I_\mathrm{sc} }{A} \right) } + A  (1+R_\mathrm{s}  G_\mathrm{sh})  \right)^2} \label{eq:1stderivativeIpvSC3D_dA}
\end{equation}
\end{proof}
\begin{lemma} \label{lemma:NRLlimit}
The maximum power point of the NRLM belongs to a curve on the I-V plane $\left(V_\mathrm{mp}^\mathrm{NRL},I_\mathrm{mp}^\mathrm{NRL}\right)$ called NRL limit. 
\end{lemma}
\begin{proof}
The NRL limit is expressed by the maximum power voltage and the maximum power current as functions of $A$ indicated by  \cref{eq:NRLM_voltage_maximum,eq:NRLM_current_maximum,eq:NRLM_voltage_maximum_auxiliar} (see details on \cref{app:nrm_limit}). Here, $V_\mathrm{mp}^{\mathrm{NRL}}\left(A\right)$ and $I_\mathrm{mp}^{\mathrm{NRL}}\left(A\right)$ represent the MPP of the NRLM-1 while $W_0(\bullet)$ indicates the zero branch of the Lambert W function. The NRL limit describes a curve on the I-V plane as shown by \cref{fig:theorem001}. This limit satisfies $\lim_{A\rightarrow0^+} \left(V_\mathrm{mp}^\mathrm{NRL},I_\mathrm{mp}^\mathrm{NRL}\right) = \left(V_\mathrm{oc},I_\mathrm{sc}\right)$ and $\lim_{A\rightarrow +\infty}\left(V_\mathrm{mp}^\mathrm{NRL},\,I_\mathrm{mp}^\mathrm{NRL}\right) = \left(\nicefrac{V_\mathrm{oc}}{2},\,\nicefrac{I_\mathrm{sc}}{2}\right)$.
\begin{align}
    z_\mathrm{mp}^\mathrm{NRL}  & \coloneqq \exp{ \left( \cfrac{V_\mathrm{oc}}{A}+1 \right) } \label{eq:NRLM_voltage_maximum_auxiliar} \\
    V_\mathrm{mp}^\mathrm{NRL}(A) & \coloneqq A  \left(W_0 \left( z_\mathrm{mp}^\mathrm{NRL} \right) - 1 \right) \label{eq:NRLM_voltage_maximum} \\
    I_\mathrm{mp}^\mathrm{NRL}(A) & \coloneqq \frac{ I_\mathrm{sc}  \exp{ \left(\cfrac{V_\mathrm{oc}}{A}\right)}  \left( W_0 \left( z_\mathrm{mp}^\mathrm{NRL} \right) - 1 \right) }{ \left( \exp{ \left( \cfrac{V_\mathrm{oc}}{A} \right) } -1 \right)  W_0 \left( z_\mathrm{mp}^\mathrm{NRL} \right) } \label{eq:NRLM_current_maximum}  
\end{align}
\end{proof}
\begin{lemma} \label{lemma:mppNSHM}
The maximum power point of the NSHM belongs to a region on the I-V plane called NSH region. This region is delimited by $I_\mathrm{pv} = I_\mathrm{sc}$, $V_\mathrm{pv} = \nicefrac{V_\mathrm{oc}}{2}$ and the NRL limit.
\end{lemma}
\begin{proof}
By extension of \cref{prop:mpp004}, the limit of the NSR region is described by all those I-V curves where $A\rightarrow0^+$, as indicated by \cref{eq:NSHM_limit002a,eq:NSHM_limit002b} (see details in \cref{appendix:mp_NSHM}). Here, the MPP voltage and MPP current are represented as functions of $R_\mathrm{s}$. It is noticed that for the limit case where $R_\mathrm{s} \rightarrow 0^+$, the MPP corresponds exactly to the NRL limit. From this representation, the boundary of the NSH region is defined by $I_\mathrm{pv} = I_\mathrm{sc}$, $V_\mathrm{pv}=\nicefrac{V_\mathrm{oc}}{2}$ and the NRL limit. \cref{fig:theorem001} shows the shape of the NSH region where the arrows on the red dashed line indicate the values of $\left(V_\mathrm{mp}^\mathrm{NSH}(R_\mathrm{s}),I_\mathrm{mp}^\mathrm{NSH}(R_\mathrm{s}) \right)$ according to $R_\mathrm{s}$.
\begin{align}
    V_\mathrm{mp}^\mathrm{NSH}(R_\mathrm{s}) \coloneqq \begin{cases}
    V_\mathrm{oc}-R_\mathrm{s} I_\mathrm{sc}, & \text{if} \quad R_{\mathrm{s}} \in \,  \biggr]  0, \cfrac{V_\mathrm{oc}}{2 I_\mathrm{sc}} \biggl[, \\
    \cfrac{V_\mathrm{oc}}{2}, & \text{if} \quad R_{\mathrm{s}} \in \biggl[ \cfrac{V_\mathrm{oc}}{2 I_\mathrm{sc}}, \cfrac{V_\mathrm{oc}}{I_\mathrm{sc}}\biggl[.
    \end{cases} \label{eq:NSHM_limit002a}
\end{align}   
\begin{align}
    I_\mathrm{mp}^\mathrm{NSH}(R_\mathrm{s})  &\coloneqq \begin{cases}
    I_\mathrm{sc}, & \text{if} \quad R_{\mathrm{s}} \in \,  \biggr]  0, \, \cfrac{V_\mathrm{oc}}{2 I_\mathrm{sc}} \biggl[, \\
    \cfrac{V_\mathrm{oc}}{2  R_\mathrm{s}} , & \text{if} \quad R_{\mathrm{s}} \in \biggl[ \cfrac{V_\mathrm{oc}}{2 I_\mathrm{sc}}, \, \cfrac{V_\mathrm{oc}}{I_\mathrm{sc}}\biggl[.
    \end{cases} \label{eq:NSHM_limit002b}
\end{align}   
\end{proof}
\begin{lemma} \label{lemma:mppNSRM}
The maximum power point of the NSRM belongs to a region on the I-V plane called NSR region. This region is delimited by $V_\mathrm{pv} = V_\mathrm{oc}$, $I_\mathrm{pv} = \nicefrac{I_\mathrm{sc}}{2}$ and the NRL limit.
\end{lemma}
\begin{proof}
By extension of \cref{prop:mpp004}, the limit of the NSR region is described by all those I-V curves where $A\rightarrow0^+$, as indicated by \cref{eq:NSRM_limit002a,eq:NSRM_limit002b} (see details in \cref{appendix:mp_NSRM}). Here, the MPP voltage and MPP current are represented as functions of $G_\mathrm{sh}$. It is noticed that for the limit case where $G_\mathrm{sh} \rightarrow 0^+$, the MPP corresponds exactly to the NRL limit. From this representation, the boundary of the NSR region is defined by $I_\mathrm{pv} = \nicefrac{I_\mathrm{sc}}{2}$, $V_\mathrm{pv}=V_\mathrm{oc}$ and the NRL limit. \cref{fig:theorem001} shows the shape of the NSR region where the arrows on the dot-dashed blue line indicate the values of $\left(V_\mathrm{mp}^\mathrm{NSR}(G_\mathrm{sh}),I_\mathrm{mp}^\mathrm{NSR}(G_\mathrm{sh}) \right)$ according to $G_\mathrm{sh}$.
\begin{align}
    V_\mathrm{mp}^\mathrm{NSR}(G_\mathrm{sh}) &\coloneqq \begin{cases}
    V_\mathrm{oc}, &\text{if} \quad G_{\mathrm{sh}} \in \biggr] 0, \, \cfrac{I_\mathrm{sc}}{2 V_\mathrm{oc}} \bigg[, \\
    \cfrac{I_\mathrm{sc}}{2 G_\mathrm{sh}},  &\text{if} \quad G_{\mathrm{sh}} \in \bigg[ \cfrac{I_\mathrm{sc}}{2 V_\mathrm{oc}},\, \cfrac{I_\mathrm{sc}}{V_\mathrm{oc}} \bigg[.
    \end{cases} \label{eq:NSRM_limit002a}
\end{align}
\begin{align}
    I_\mathrm{mp}^\mathrm{NSR}(G_\mathrm{sh}) &\coloneqq \begin{cases}
    I_\mathrm{sc}-G_\mathrm{sh}  V_\mathrm{oc}, &\text{if} \quad G_{\mathrm{sh}} \in \biggr] 0, \, \cfrac{I_\mathrm{sc}}{2 V_\mathrm{oc}} \bigg[, \\
    \cfrac{I_\mathrm{sc}}{2} , &\text{if} \quad G_{\mathrm{sh}} \in \bigg[ \cfrac{I_\mathrm{sc}}{2 V_\mathrm{oc}},\, \cfrac{I_\mathrm{sc}}{V_\mathrm{oc}} \bigg[.
    \end{cases} \label{eq:NSRM_limit002b}
\end{align}
\end{proof}
\begin{theorem} \label{theorem:mppSDM}
The maximum power point of the SDM is limited to a region on the I-V plane called SDM region. This region is defined by the union of the NSH region and the NSR region.
\end{theorem}
\begin{proof}
\par Let be a SDM-3 with $V_\mathrm{oc}$ and $I_\mathrm{sc}$ given. Let splite the region $\nicefrac{V_\mathrm{oc}}{2}<V_\mathrm{mp}<V_\mathrm{oc}$ and $\nicefrac{I_\mathrm{sc}}{2}<I_\mathrm{mp}<I_\mathrm{sc}$ by means of the NRL limit. Then, there are two possible regions where the MPP of the SDM can exist. By the Kirchhoff Current Law (KCL) and the Kirchhoff Voltage Law (KVL), it is accomplished that the output voltage of the NSHM-2 is lower than the NRLM-1 by a factor $R_\mathrm{s}  I_\mathrm{pv}$ for $I_\mathrm{pv} \in \left]0,I_\mathrm{sc}\right[$. Then, since the deviation on the voltage depends exclusively on $R_\mathrm{s}$, one of the regions must correspond to the NSH region. Similarly, the output current of NSRM-2 is lower than the NRLM-1 by a factor $G_\mathrm{sh}  V_\mathrm{pv}$ for $V_\mathrm{pv} \in \left]0,\,V_\mathrm{oc}\right[$. Then, since the deviation on the current depends exclusively on $G_\mathrm{sh}$, one of the regions must correspond to the NSR region. Therefore, the SDM region can be described by the union of the NSH region and the NSR region. This situation is presented in \cref{fig:theorem001}. Here, the theoretical region is bounded by the union of the dashed red line and the dot-dashed blue line.
\end{proof}

\section{Implicit SDM representation} \label{sec:feasible_domain}
In this section the SDM is expressed implicitly as a function of a single parameter. Then, it is possible to define the set of SDM  intersecting the cardinal points such that $\Theta \in \mathbb{R}_+^5$ (\emph{cf.} \cref{theorem_one-dimensional}). For this, the SDM is represented as a function of the equivalent factor of the diode and the series resistance (\emph{cf.} \cref{ssec:two_parameters_sdm_representation}). 
\subsection{Two-parameters SDM representation} \label{ssec:two_parameters_sdm_representation}
Consider three points on the I-V curve: $(0,I_\mathrm{sc})$, $(V_\mathrm{oc},0)$ and any point $(V_\mathrm{x},I_\mathrm{x})$ selected according to \cref{lemma:concavity} such that $0<V_\mathrm{x}<V_\mathrm{oc}$ and $I_\mathrm{sc}(1- \nicefrac{V_\mathrm{x} }{V_\mathrm{oc}})<I_\mathrm{x}<I_\mathrm{sc}<I_\mathrm{x}<I_\mathrm{sc}$. Then $I_\mathrm{ph}$, $I_\mathrm{o}$, and $G_\mathrm{sh}$ can be expressed as functions of $\Theta_2 \coloneqq \{ A, R_\mathrm{s} \}$. Replacing $I_\mathrm{ph}(\Theta_2)$, $I_\mathrm{o}(\Theta_2)$, and $G_\mathrm{sh}(\Theta_2)$ on \cref{def:SDM}, the two-parameters SDM (SDM-2) is formally defined as indicated \cref{def:2P_SDM}. Here, $F_\mathrm{SD2}(X;\Theta_2)$ is presented by \cref{eq:2D_sdm_Fsd2} (see details in \cref{app:two-parameter_SDM}).
%
%
\begin{multline}
\label{def:2P_SDM}  
\text{SDM-2} \coloneqq \{ (X,\Theta_2 ) \in \mathbb{R}^2 \times \mathbb{R}_+^2 \mid \\ F_\mathrm{SD2}(X;\Theta_2) = 0,\, \{ I_\mathrm{ph}(\Theta_2),\, I_\mathrm{o}(\Theta_2),\ G_\mathrm{sh}(\Theta_2) \} >0\} 
\end{multline}
The SDM-2 ensures $I_\mathrm{ph}(\Theta_2),I_\mathrm{o}(\Theta_2)>0$ if $G_\mathrm{sh}(\Theta_2)>0$ as stated in \cref{prop:2D_restriction}. Then, from \cref{prop:2D_restriction}, there exists a series resistance $R_\mathrm{s}^\mathrm{sh}$ as a function of $A$ such that $G_\mathrm{sh}(\Theta_2)=0$, demonstrated in \cref{prop:2D_Rsi(A)}. This function represents the maximum possible value for $R_\mathrm{s}$ given $A$, as illustrated in \cref{fig:theorem002}. Here, it is appreciated that due to the fact that $R_\mathrm{s}^\mathrm{sh}(A)$ is a decreasing function, there exists a maximum value $A_\mathrm{max}^\mathrm{sh}$ such that $R_\mathrm{s}=0$. 
\begin{proposition} \label{prop:2D_restriction}
The SDM-2 such that $G_\mathrm{sh}(\Theta_2)>0$ implies $\{ I_\mathrm{ph}(\Theta_2),\,I_\mathrm{o}(\Theta_2)\}>0$.    
\end{proposition}
\begin{proof}
\par Let express $I_\mathrm{ph}(\Theta_2)$, $I_\mathrm{o}(\Theta_2)$ and $G_\mathrm{sh}(\Theta_2)$ as functions of $F_\mathrm{sh}(\Theta_2)$ and $F_\mathrm{cd}(\Theta_2)$ (see details in \cref{app:two-parameter_SDM}). Let suppose that $F_\mathrm{sh}(\Theta_2)<0$. Then, $G_\mathrm{sh}(\Theta_2)>0$ implies $F_\mathrm{cd}(\Theta_2)<0$. However, by \cref{lemma:concavity}, $I_\mathrm{o}(\Theta_2)>0$ requires $F_\mathrm{cd}(\Theta_2)>0$. Therefore, by contradiction, $F_\mathrm{sh}(\Theta_2)$ must be positive. Then, since $F_\mathrm{sh}(\Theta_2)>0$ ensures $I_\mathrm{ph}(\Theta_2)>0$, it is said that $G_\mathrm{sh}(\Theta_2)>0$ implies $\{ I_\mathrm{ph}(\Theta_2),\,I_\mathrm{o}(\Theta_2)\}>0$.
\end{proof}
\begin{lemma} \label{prop:2D_Rsi(A)}
There exists a decreasing series resistance $R_\mathrm{s}^\mathrm{sh}$ as a function of $A$ such that $G_\mathrm{sh}(\Theta_2)=0$ and $\Theta_2 \in \mathbb{R}_+^2$.
\end{lemma}
\begin{proof}
\par To demonstrate the existence of $R_\mathrm{s}^\mathrm{sh}(A)$, the implicit function theorem is applied as indicated by \cref{eq:implicit_theorem}. Let's express $G_\mathrm{sh}(\Theta_2)$ as $\nicefrac{F_\mathrm{sh}(\Theta_2)}{F_\mathrm{cd}(\Theta_2)}$. Since it must be analyzed the tuple $\Theta_2$ such that $G_\mathrm{sh}(\Theta_2)=0$, the problem is equivalent to study $F_\mathrm{sh}(\Theta_2)=0$. Then,  $\odif[switch-*=true]{A}{R_\mathrm{s}^\mathrm{sh}}$ exist if and only if $\partial_{R_\mathrm{s}}  F_\mathrm{sh}(\Theta_2) \neq 0$. This is verified by \cref{eq:implicit_theorem02a}. Here, since $R_\mathrm{s}<\nicefrac{V_\mathrm{oc}}{I_\mathrm{sc}}$, it is ensured $R_\mathrm{s}  I_\mathrm{sc} < V_\mathrm{x} + R_\mathrm{s}  I_\mathrm{x} < V_\mathrm{oc} $ for $ 0<V_\mathrm{x}<V_\mathrm{oc}$ and $I_\mathrm{sc}(1- \nicefrac{V_\mathrm{x} }{V_\mathrm{oc}})<I_\mathrm{x}<I_\mathrm{sc}$ (see details in \cref{app:two-parameter_SDM}). Then, $\partial_{R_\mathrm{s}} F_\mathrm{sh}(\Theta_2)<0$ and there must exist a function $R_\mathrm{s}^\mathrm{sh}(A)$.
\begin{equation}
    \label{eq:implicit_theorem}
    \odif[switch-*=true]{A}{R_\mathrm{s}^\mathrm{sh}} \coloneqq - \cfrac{\partial_{A} F_\mathrm{sh}(\Theta_2)}{\partial_{R_\mathrm{s}} F_\mathrm{sh}(\Theta_2)}
\end{equation}
\begin{equation}
    \partial_{R_\mathrm{s}} F_\mathrm{sh} (\Theta_2) \coloneqq - \cfrac{I_\mathrm{sc}  I_\mathrm{x}}{A}  \left( \exp{\left( \cfrac{V_\mathrm{x}+R_\mathrm{s}  I_\mathrm{x}}{A}\right)} \right.  \left. -  \exp{\left( \cfrac{R_\mathrm{s}  I_\mathrm{sc}}{A} \right)} \right) \label{eq:implicit_theorem02a}
\end{equation}
\par To check the sign of $\odif[switch-*=true]{A}{R_\mathrm{s}^\mathrm{sh}}$ is necessary to analyze $\partial_{A} F_\mathrm{sh}(\Theta_2)$, indicated in \cref{eq:implicit_theorem03a}. In this sense, for $A\rightarrow 0^+$ it is satisfied $ F_\mathrm{sh}(\Theta_2) > 0$ and $\partial_A F_\mathrm{sh}(\Theta_2) < 0$. On the other hand, for $A\rightarrow +\infty$ it is satisfied $ F_\mathrm{sh}(\Theta_2)<0$ and $\partial_A F_\mathrm{sh}(\Theta_2) > 0$. Then, considering that there exist only one tuple $\Theta_2$ such that $ F_\mathrm{sh}(\Theta_2)=0$, it is possible to ensure that $\partial_A F_\mathrm{sh}(\Theta_2) < 0$ in that point. Therefore, $R_\mathrm{s}^\mathrm{sh}(A)$ must be a decreasing function.
\begin{multline}
    \partial_{A} F_\mathrm{sh}(\Theta_2) \coloneqq \frac{1}{A^2}  \Bigg( - V_\mathrm{oc}  (I_\mathrm{sc}-I_\mathrm{x})  \exp{\left( \cfrac{V_\mathrm{oc}}{A} \right)} \\ + I_\mathrm{sc}  (V_\mathrm{x} + R_\mathrm{s}  I_\mathrm{x})  \exp{\left( \cfrac{V_\mathrm{x}+R_\mathrm{s} I_\mathrm{x}}{A} \right)} \\ - R_\mathrm{s}  I_\mathrm{sc}  I_\mathrm{x}  \exp{\left( \cfrac{R_\mathrm{s} I_\mathrm{sc}}{A} \right)} \Bigg)     \label{eq:implicit_theorem03a}
\end{multline}
\end{proof} 
\subsection{Implicit approach} \label{ssec:implicit_approach}
The SDM can be expressed implicitly as a function of $A$ or $R_\mathrm{s}$ if the point $(V_\mathrm{x},I_\mathrm{x})$ corresponds to the MPP. For this, the derivative of the SDM-2 according to the voltage $\partial_{V_\mathrm{pv}} F_\mathrm{SD2}(X;\Theta_2)$ has to be calculated. Then, evaluating $\partial_{V_\mathrm{pv}} F_\mathrm{SD2}(X;\Theta_2)$ at the MPP and applying \cref{lemma:photovoltaic_mpp}, the maximum power point function $F_\mathrm{mp}(\Theta_2)$ indicated by \cref{eq:fmp_theta2} is obtained (see details in \cref{app:mpp_sdm-2}). Therefore, the SDM can be represented implicitly as a function of $A$ or $R_\mathrm{s}$, denominated as one-parameter SDM (SDM-1). For instance, \cref{def:1P_SDM} presents the formal definition for the SDM-1 as a function of $A$. Then, under this definition, \cref{theorem_one-dimensional} defines the complete family of SDM-1 intersecting the cardinal points.
%
\begin{multline}
\label{def:1P_SDM}
\text{SDM-1} \coloneqq \{ (X,A) \in \mathbb{R}^2 \times \mathbb{R}_+ \mid \\ \{ F_\mathrm{SDM-2}(X;\Theta_2), \, F_\mathrm{mp}(\Theta_2) \} = 0,\, G_\mathrm{sh}(\Theta_2)>0 \} 
\end{multline}
\cref{fig:theorem002} illustrates the behavior of $R_\mathrm{s}^\mathrm{mp}(A)$ for two $(V_\mathrm{mp},I_\mathrm{mp})$ tuples. \cref{fig:theorem002a} shows $R_\mathrm{s}^\mathrm{mp}(A)$ calculated with a $(V_\mathrm{mp},I_\mathrm{mp})$ tuple located in the NSH region. Here, $R_\mathrm{s}^\mathrm{mp}(A)$ intersects $R_\mathrm{s}^\mathrm{sh}(A)$ for an $R_\mathrm{s,\,min}$ value greater than zero. On the other hand, \cref{fig:theorem002b} shows $R_\mathrm{s}^\mathrm{mp}(A)$ calculated with a $(V_\mathrm{mp},I_\mathrm{mp})$ tuple located in the NSR region. For this case, $R_\mathrm{s}^\mathrm{mp}(A)$ intersects $R_\mathrm{s}^\mathrm{sh}(A)$ at $R_\mathrm{s,\,min}$ equal to zero. Here, it is shown graphically that $R_\mathrm{s,\,min} \geqslant 0$ depending on the values of $(V_\mathrm{mp},\,I_\mathrm{mp})$.

By \cref{theorem:mppSDM,theorem_one-dimensional} it is extracted that depending on the location of $(V_\mathrm{mp},\,I_\mathrm{mp})$, the solution $(A_\mathrm{max},R_\mathrm{s,\,min})$ must be calculated accordingly. In this sense, if $(V_\mathrm{mp},\,I_\mathrm{mp})$ belongs to the NSH region, then $A_\mathrm{max}$ is a solution of $F_\mathrm{mp}(A_\mathrm{max},\,R_\mathrm{s,\,min}) = 0$ and $ F_\mathrm{sh}(A_\mathrm{max},\,R_\mathrm{s,\,min}) = 0$. On the other hand, if $(V_\mathrm{mp},\,I_\mathrm{mp})$ belongs to the NSR region, then $R_\mathrm{s,\,min} = 0$ and $A_\mathrm{max}$ is the solution of $F_\mathrm{mp}(A_\mathrm{max},\,0) = 0$. 
\begin{theorem} \label{theorem_one-dimensional}
The set of SDM satisfying the cardinal points is defined implicitly by the series resistance $R_\mathrm{s}^\mathrm{mp}$ as a function of $A$ such that $ G_\mathrm{sh}(\Theta_2)>0$ and $\Theta_2 \in \mathbb{R}_+^2$.
\end{theorem}
\begin{proof}
To demonstrate the existence of $R_\mathrm{s}^\mathrm{mp}(A)$, the implicit function theorem is applied as indicated by \cref{eq:imp_theorem_2a}. Then, $\odif[switch-*=true]{A}{R_\mathrm{s}^\mathrm{mp}}$ exist if and only if $\partial_{R_\mathrm{s}} F_\mathrm{mp} (\Theta_2) \neq 0$ such that $F_\mathrm{mp}(\Theta_2)=0$. This can be verified by means of \cref{eq:imp_theorem_2b,eq:imp_theorem_2b1,eq:imp_theorem_2b2}. Let $R_\mathrm{s} \in \left] 0, \nicefrac{\left( V_\mathrm{oc} - V_\mathrm{mp} \right)}{I_\mathrm{sc}} \right[$ and a set of cardinal points according to \cref{prop:minimum_power} be given. Then, for $A \rightarrow 0^+$, $\partial_{R_\mathrm{s}} F_\mathrm{mp} (\Theta_2) \rightarrow +\infty$. On the other hand, if  $A \rightarrow +\infty$, there are two possibles situations for $\partial_{R_\mathrm{s}} F_\mathrm{mp} (\Theta_2)$: (i) it tends to $0^+$ and $R_\mathrm{s}^{\mathrm{mp}}(A)$ exist for all $A$ or (ii) it tends to $0^-$ and $R_\mathrm{s}^{\mathrm{mp}}(A)$ is discontinuous for one given $A$ value. In this sense, $R_\mathrm{s}^{\mathrm{mp}}(A)$ exist for $G_\mathrm{sh}(\Theta_2)>0$ and $\Theta_2 \in \mathbb{R}_+^2$, except for one and only one $A \in \left] 0, +\infty \right[$.
\begin{align}
    \label{eq:imp_theorem_2a} 
    \odif[switch-*=true]{A}{R_\mathrm{s}^\mathrm{mp}}\coloneqq &- \cfrac{\partial_{A} F_\mathrm{mp}(\Theta_2)}{\partial_{R_\mathrm{s}} F_\mathrm{mp}(\Theta_2)} 
\end{align}
\begin{multline}
    \label{eq:imp_theorem_2b}   
    \partial_{R_\mathrm{s}} F_\mathrm{mp}(\Theta_2) \coloneqq I_\mathrm{mp} X_\mathrm{2}  \exp{\left( \cfrac{R_\mathrm{s} I_\mathrm{sc}}{A} \right)}  \\ I_\mathrm{mp} \Bigg( \frac{
            X_1  (V_\mathrm{mp}-  R_\mathrm{s}  I_\mathrm{mp})
        }{A} - X_\mathrm{2} \Bigg)  \exp{\left( \cfrac{V_\mathrm{mp}+R_\mathrm{s} I_\mathrm{mp}}{A} \right)} 
\end{multline}
\begin{align}
    \label{eq:imp_theorem_2b1}   
    X_1 &\coloneqq V_\mathrm{oc}  I_\mathrm{mp} + V_\mathrm{mp}  I_\mathrm{sc}  - V_\mathrm{oc}  I_\mathrm{sc} \\
    \label{eq:imp_theorem_2b2}   
    X_2 &\coloneqq I_\mathrm{sc}(2V_\mathrm{mp}-V_\mathrm{oc})
\end{align}
%

\begin{figure}[!ht]
\centering
\begin{subfigure}{0.49\textwidth}
    \includegraphics[width=\textwidth]{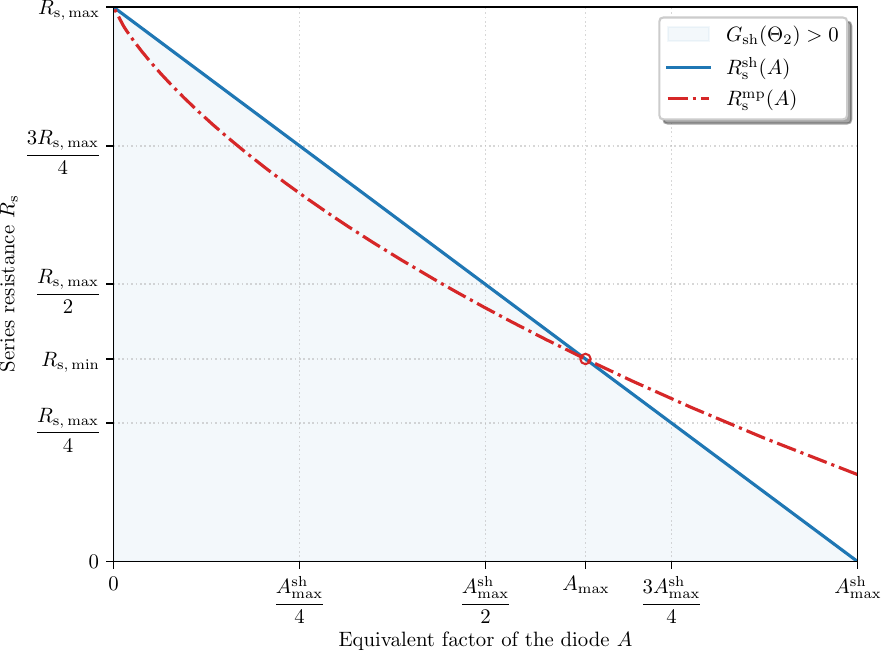}
    \caption{ }
    \label{fig:theorem002a}
\end{subfigure}
\hfill
\begin{subfigure}{0.49\textwidth}
    \includegraphics[width=\textwidth]{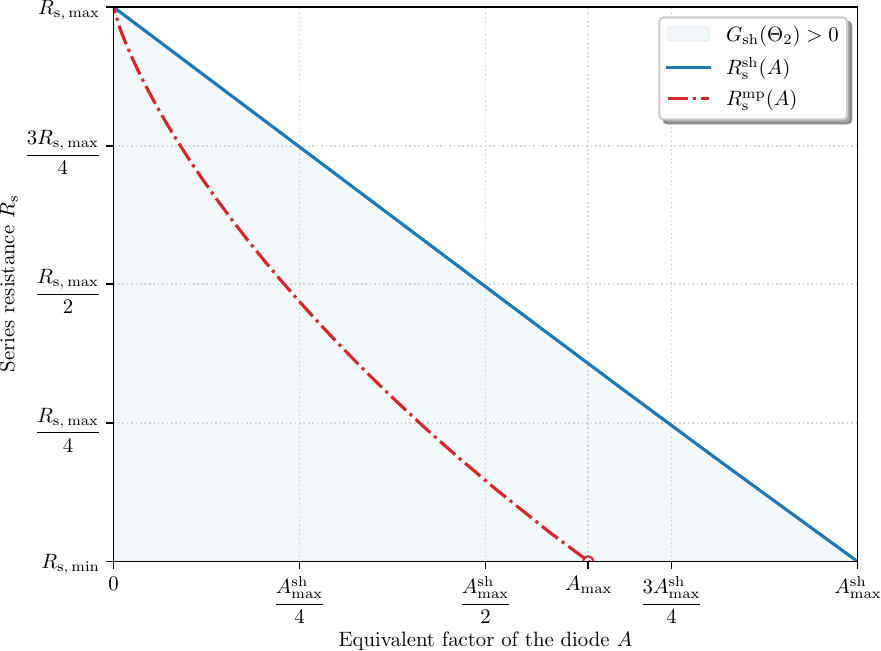}
    \caption{ }
    \label{fig:theorem002b}
\end{subfigure}
\caption{Representation of $R_\mathrm{s}^\mathrm{mp}(A)$ and $R_\mathrm{s}^\mathrm{sh}(A)$. The zone where $G_\mathrm{sh}(\Theta_2)>0$ is indicated in blue. Panel (a) shows the case for an MPP in the NSH region, where $R_\mathrm{s,\,min}>0$. Panel (b) shows the case for an MPP in the NSR region, where $R_\mathrm{s,\,min}=0$.}
\label{fig:theorem002}
\end{figure}
\end{proof}

\section{General methodology framework}
\label{sec:methodology_description}

Let $\fpv(\Vpv,\,\Ipv;\,\tpv)$ be a differentiable function relating the photovoltaic current $\Ipv$ and the photovoltaic voltage $\Vpv$ through the set of $k$ parameters $\tpv$ such that $\tpv \in \R^k_+$.
Usually, the function $\fpv$ is non-linear having one or more exponential terms, as in the case of the single-diode model or the double-diode model.
Therefore, the identification of the set of parameters $\tpv$ result in an extensive process.
To simplify this task, consider the incorporation of the three cardinal points in the identification process: short-circuit point $(0,\,\Isc)$, open-circuit point $(\Voc,\,0)$, and the maximum power point $(\Vmp,\,\Imp)$.
Then, three constraints are incorporated:
\begin{align}
    \label{constrain_01}
    \fpv(0,\,\Isc;\,\tpv) &= 0, \\
    \label{constrain_02}
    \fpv(\Vmp,\,\Imp;\,\tpv) &= 0, \\
    \label{constrain_03}
    \fpv(\Voc,\,0;\,\tpv) &= 0.
\end{align}
Moreover, from the maximum power point, a fourth constraint is included:
\begin{equation}
    \label{constrain_04}
    \Imp \left( \pdif[switch-*=true]{\Ipv}{\fpv} \right) \big|_\mathrm{mp} - \Vmp \left( \pdif[switch-*=true]{\Vpv}{\fpv} \right) \big|_\mathrm{mp} = 0.
\end{equation}
In this sense, any set of parameters $\tpv \in \R^k_+$ such that \cref{constrain_01,constrain_02,constrain_03,constrain_04} are satisfied, is a solution.

\section{Computation algorithm} \label{sec:comp_alg}
\begin{figure}[!hbt]
\centering
\includegraphics[width=\textwidth]{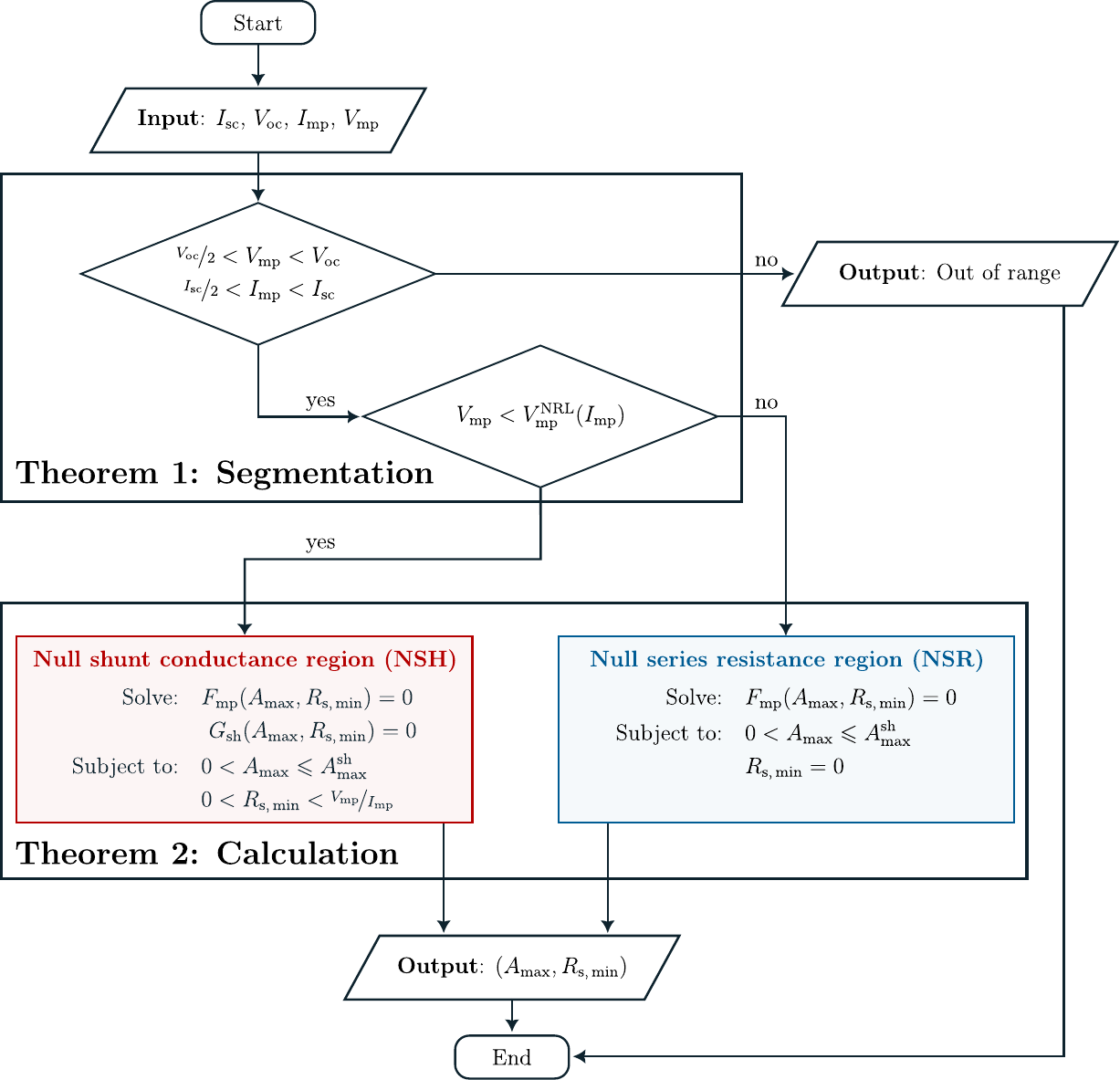}
\caption{Flowchart to calculate $A_{\mathrm{max}}$ and $R_\mathrm{s,\,min}$. The application of \cref{theorem:mppSDM} segments the input data. Next, applying \cref{theorem_one-dimensional}, it is possible to calculate $A_{\mathrm{max}}$ and $R_\mathrm{s,\,min}$.}
\label{fig:flow_diagram}
\end{figure}

In this section, it is presented the main steps to calculate  $A_\mathrm{max}$ and $R_\mathrm{s,\,min}$ employing \cref{theorem:mppSDM,theorem_one-dimensional}. \cref{fig:flow_diagram} presents the flow chart of the computation algorithm where two main steps are identified: (i) segmentation and (ii) calculation. The segmentation step is performed by \cref{theorem:mppSDM} and locates the cardinal points in the plane shown by \cref{fig:theorem001}. In this sense, the input data may belong to the NSHM, the NSRM, or be out of range. To determine this, $V_\mathrm{mp}^\mathrm{RL}$ is expressed as a function of $I_\mathrm{mp}$ (see details in \cref{app:nrm_limit}). Then, if $\nicefrac{V_\mathrm{oc}}{2}<V_\mathrm{mp}<V_\mathrm{mp}^\mathrm{RL}$, the MPP of the input data belongs to the NSHM. Otherwise, if $V_\mathrm{mp}^\mathrm{RL} \leqslant V_\mathrm{mp}<V_\mathrm{oc}$ the MPP is located on the NSRM. The second step is related to the calculation of $A_\mathrm{max}$ and $R_\mathrm{s,\,min}$ based on \cref{theorem_one-dimensional}. Here, depending on the location of the cardinal points on \cref{fig:theorem001}, there are two possible ways to calculate $A_\mathrm{max}$ and $R_\mathrm{s,\,min}$.

The Levenberg-Marquardt algorithm is used to solve the equation system. However, to operate, this algorithm requires an initial seed. For this problem, the initial seeds are selected based on whether the MPP belongs to the NSR region or the NSH region. In this sense, for the case where MPP belongs to the NSR region, the initial value $A_\mathrm{init}$ is selected such that $F_\mathrm{mp}(A_\mathrm{init},0)<0$. On the other hand, if MPP belongs to the NSH region, $A_\mathrm{init}$ is such that $R_\mathrm{s}^\mathrm{sh}(A_\mathrm{init})>R_\mathrm{s}^\mathrm{mp}(A_\mathrm{init})$. The initial value for the series resistance $R_\mathrm{s,\,init}$ is assumed to be equal to $R_\mathrm{s}^\mathrm{sh}(A_\mathrm{init})$. 

\section{Results and discussion} \label{sec:results}
    \subsection{Analysis on synthetic data} \label{sec:results_scaled}
        To test the calculation algorithm, a regular grid of tuples $(v_\mathrm{mp},\,i_\mathrm{mp})$ is generated. Here, every tuple represents the scaled version of the maximum power voltage and current (see details in \cref{app:scaled}). In this test, $v_\mathrm{mp}$ and $i_\mathrm{mp}$ are represented as two equal vectors starting from $\nicefrac{1}{2}$ generating a grid with a total of $40\,000$ data points. Then, by \cref{theorem:mppSDM}, the grid points are divided as follows: (i) $23\,494$ on the NSR, (ii) $14\,558$ on the NSH, and (iii) $1\,948$ points that cannot be processed due to computational limitations.
        
\begin{figure}[!ht]
        \centering
        \begin{subfigure}{0.49\textwidth}
            \includegraphics[width=\textwidth]{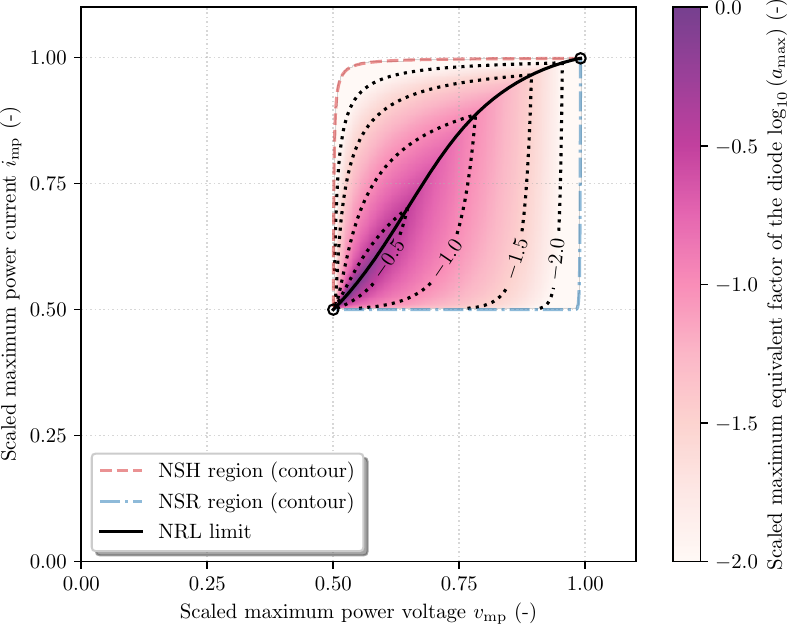}
            \caption{ } \label{fig:results_001a}
        \end{subfigure}
        \hfill
        \begin{subfigure}{0.49\textwidth}
            \includegraphics[width=\textwidth]{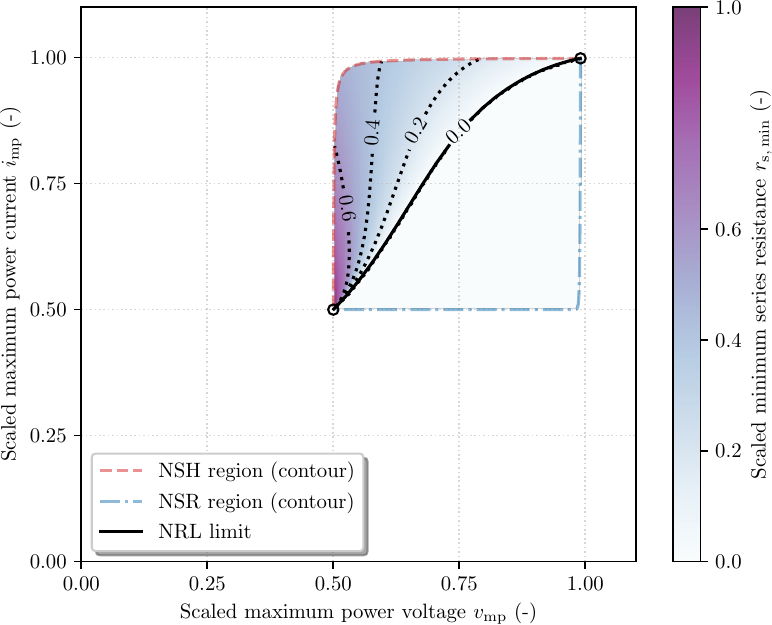}
            \caption{ } \label{fig:results_001b}
        \end{subfigure}
        \caption{Calculated $a_\mathrm{max}$ and $r_\mathrm{s,\,min}$ from the artificial data. The contour of the NSH region (red zone) and the contour of the NSR region (blue zone) are included. Panel (a) shows the behavior of $a_\mathrm{max}$ according to $(v_\mathrm{mp},\,i_\mathrm{mp})$. Panel (b) presents the behavior of  $r_\mathrm{s,\,min}$ and $(v_\mathrm{mp},\,i_\mathrm{mp})$.}
        \label{fig:results_001}
        \end{figure}

        \cref{fig:results_001} presents the calculated scaled parameters $a_\mathrm{max}$ and $r_\mathrm{s,\,min}$ as functions of $(v_\mathrm{mp},\,i_\mathrm{mp})$. From \cref{fig:results_001a} it is appreciated that $a_\mathrm{max}$ is infinite for $(v_\mathrm{mp},\,i_\mathrm{mp})  \rightarrow (\nicefrac{1}{2},\,\nicefrac{1}{2})$, indicating that no matter what $A$ value is selected there is always a SDM-1 matching the cardinal points. Then, as $(v_\mathrm{mp},\,i_\mathrm{mp})$ increases, $a_\mathrm{max}$ tends to decrease, being minimum at $(v_\mathrm{mp},\,i_\mathrm{mp}) \rightarrow (1,1)$. \cref{fig:results_001b} presents the calculated values for $r_\mathrm{s,\,min}$. It is noted that $r_\mathrm{s,\,min}$ remains finite and lower than 1 for all cases. However, for the limit case where $(v_\mathrm{mp},\,i_\mathrm{mp}) \rightarrow (\nicefrac{1}{2},\,\nicefrac{1}{2})$, $r_\mathrm{s,\,min} = r_\mathrm{s,\,max} = 1$, therefore, $a$ can take any value, as explained above. On the other hand, if $(v_\mathrm{mp},\,i_\mathrm{mp}) \rightarrow (1,1)$, $r_\mathrm{s,\,min}$ is equal to zero and the solar model is represented by an ideal current source. 

        \begin{remark}
            \label{remark01}
             From \cref{fig:results_001a}, it is followed that $a_\mathrm{max}\rightarrow0^+$ in the vicinity of the contour of the SDM region. However, for such low values of $a_\mathrm{max}$ is not possible to compute the contour of the SDM region. Therefore, there must exist an infimum value $A_\mathrm{inf}$ defined by the maximum float value supported by the computation program $N_\mathrm{float}^\mathrm{max}$. For the presented methodology, $A_\mathrm{inf}$ is defined as indicated by \cref{eq:infimum_a}. In this way, it is ensured the computation of the contour of the SDM region.
            \begin{equation}
                \label{eq:infimum_a}
                A_\mathrm{inf} \coloneqq \cfrac{V_\mathrm{oc}}{\ln{(N_\mathrm{float}^\mathrm{max})}} 
            \end{equation}
        \end{remark}

    \subsection{Application on the CEC database} \label{sec:results_cec}
        To test the calculation algorithm on real data, we used the cardinal points at the Standard Test Conditions ($1\,000$ Wm$^{-2}$, $25$ °C) from the California Energy Commission database.  To date 25/05/2023, this database counts on $17\,711$ data sheets samples of different modules technologies. However, this database presents several repeated data in terms of cardinal points, $N_\mathrm{s}$, $N_\mathrm{p}$ and technology. Therefore, after filtering the original database, the new amount of data equals $8\,600$ samples distributed: Mono crystalline-Silicon (c-Si) ($5\,542$ samples -- $64.44$ \%), Multi c-Si ($2\,832$ samples -- $32.93$ \%), Thin Film ($180$ samples -- $2.09$ \%), Cadmium Telluride (CdTe) ($30$ samples -- $0.35$ \%) and Copper-Indium-Gallium-Selenide (CIGS) ($16$ -- $0.19$ \%). To represent the filtered database, the scaling procedure indicated in \cref{app:scaled} is adopted. \cref{fig:results_001} presents the scaled maximum power point of the data set classified by the different modules technologies. As a reference, the NSH region and the NSR region are included. It is noted that all data are contained in the MPP contour except for two $(v_\mathrm{mp},\,i_\mathrm{mp})$ pairs, which are not considered in the analysis. \par
        
        \begin{figure}[!hbt]
                \centering
                \includegraphics[width=0.75\textwidth]{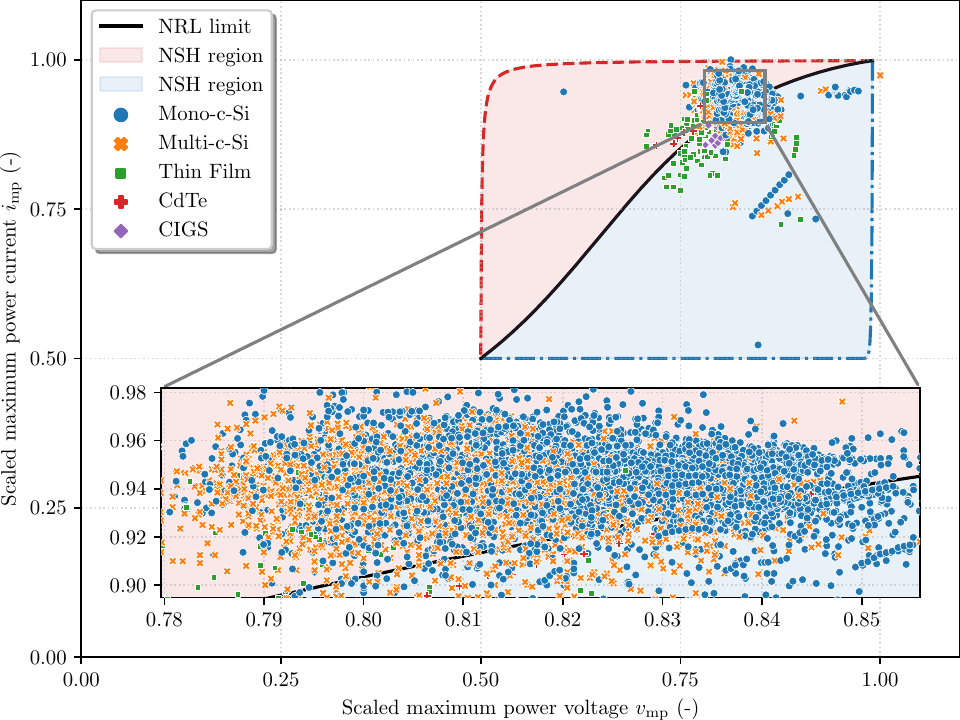}
                \caption{Scaled maximum power point $(v_{\mathrm{mp}},\,i_{\mathrm{mp}})$ corresponding to the CEC database classified by technology. The NSH region (red zone), the NSR region (blue zone), and the NRL limit (black line) are included. The limits of the zoomed region are selected so that the minimum and maximum values are within two standard deviations of the mean value of the current and voltage, as corresponding. } \label{fig:results_003}
        \end{figure}
\begin{figure}[!ht]
            \centering
            \includegraphics[width=0.75\textwidth]{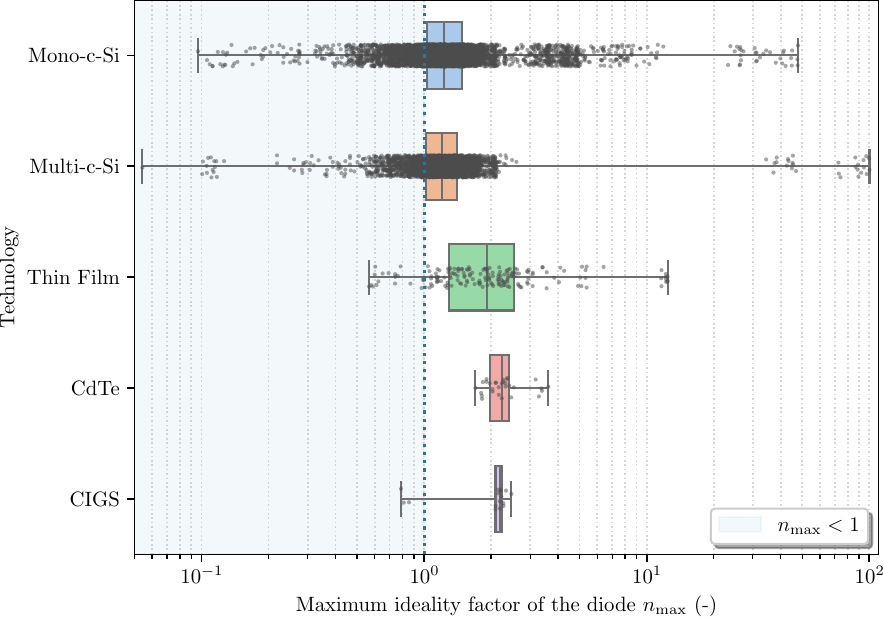}
            \caption{ Box plot of $n_\mathrm{max}$ from the filtered CEC database classified by technology. The region where $n_\mathrm{max}<1$ is indicated in red. } 
            \label{fig:results_004}
        \end{figure}

        %
            %
        After processing the CEC database, it is found that the data is segmented as follows: (i) $1\,294$ samples on the NSR region ($15.05$ \%) and (ii) $7\,304$ samples on the NSH region ($84.95$ \%). The execution time for extracting the boundaries for all $8\,600$ non-repeated samples was performed on the code ocean virtual machine in around $20$ seconds. From this result, since $N_\mathrm{s}$ is available, the ideality factor of the diode $n$ can be calculated as indicated by \cref{eq:ideality}. \cref{fig:results_004} shows the calculated ideality factor of the diode. It is observed that $n_\mathrm{max}$ for mono-c-Si and multi-c-Si technologies ranges approximately from $10^{-1}$ to $10^{2}$, being concentrated around $1$ and $2$.
        
        The calculated values for $n_\mathrm{max}$ are similar to the presented in \cite{toledo_-depth_2021,laudani_identification_2014}. In this sense, most of the $n_\mathrm{max}$ values for mono c-Si and multi c-Si are located between $1$ and $2$ ($5\,944$ samples), which has a physical meaning. However, there are $1\,883$ samples where $n_\mathrm{max}$ is less than $1$ (blue colored region), indicating that the SDM is not suitable for modeling those solar panels. Regarding the cases where $n_\mathrm{max}$ is higher than $2$, there exist solutions in the range $]1,\,2[$, implying that it is possible to calculate a SDM with physical meaning. About the Thin Film, CdTe, and CIGS technologies, it is observed that $n_\mathrm{max}$ tends to be higher than $1$. However, the data set is not large enough to draw a conclusion on these technologies. 

    \subsection{Identification of the optimal SDM}
        \label{sec:results_nrel}

        
        In the present section, our goal is to identify the optimal $A$ value for a measured I-V curve given an objective function. Then, the best set of parameters $\widehat{\Theta}$ is determined from $A$. The objective function to be minimized corresponds to the scaled mean signed deviation (S-MSD), calculated as indicated by \cref{eq:s-msd}. Here, the function $f_\mathrm{sd}(\boldsymbol{x};\,\boldsymbol{\theta_\mathrm{sd}})$ is used as basis, as proposed by Oliva \textit{et al.} \cite{oliva_review_2019}. The optimization is performed employing the LMFIT python package \cite{newville_lmfit_2014}. Then, the quality of the identified model is evaluated according to S-MSD. 
        

            \begin{equation}
                \label{eq:s-msd}
                \text{S-MSD}(X_\mathrm{meas};\,\Theta) \coloneqq \cfrac{1}{I_\mathrm{sc} n_\mathrm{data}} \sum_{i=1}^{n_\mathrm{data}} F_\mathrm{SD}(X_\mathrm{meas};\,\Theta)
            \end{equation}
        
        Tests are performed on a public database provided by the National Renewable Energy Laboratory (NREL) \cite{marion_users_2014}. 
        The data used in this study consist on $38\,929$ I-V curves measured on a mSi460A8 module between January 21, 2011 and March 4, 2012 in Cocoa (Florida). These data have been selected because a polycrystalline silicon module is used, which is very common in the industry \cite{iea-pvps_photovoltaic_2022}. 
        
\begin{figure}[!ht]
            \centering
            \includegraphics[width=0.75\textwidth]{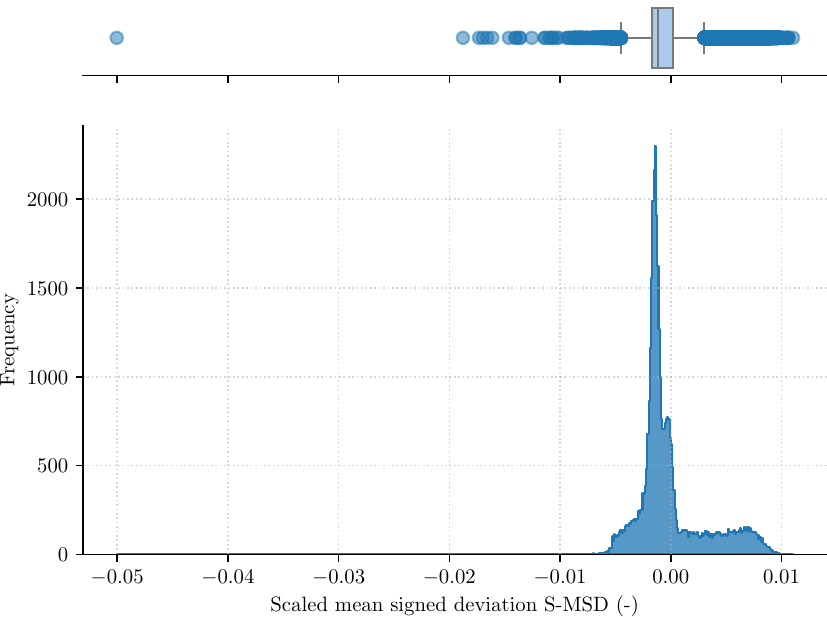}
            \caption{Histogram of the S-MSD determined for the mSi460A8 module located at Cocoa, Florida. At the top, the boxplot of the data is added to indicate the presence of outliers.} 
            \label{fig:nrel_histogram}
        \end{figure}

        The result of the processed dataset is summarized in the histogram depicted by \cref{fig:nrel_histogram}. On the top of the graphic, the boxplot of the data is included to show the presence of outliers and the quartiles 1, 2, and 3. Statistical data and information of the displayed I-V curve are incorporated in \cref{app:nrel}. The precision of this method is dependent on the estimation of the cardinal points and the quality of the measured data. However, since only one parameter is identified, the presented computation methodology compared to methods optimizing more than 1 parameter should be less computer intensive and, therefore, faster.

    \begin{remark}
        In some practical cases, the cardinal points from a measured I-V curve do not lie in the SDM region. This may result from non-uniformities of the curve or a poor estimation of the cardinal points.
        Therefore, to find the best SDM for these cases, one or several conditions (SCP, OCP or MPP) must be relaxed. 
        Naturally, such a relaxation leads to reaching a model that does not satisfy all the initial estimations of the cardinal points.

        Additionally, for the cases where the MPP is close (or outside) to the boundaries of the SDM region, it is not possible to satisfy simultaneously the cardinal points
        (see Remark 1).
        For this, the use of a piece-wise model is proposed (see \cref{eq:piecewise_current_A0}).
        Here, estimations for the maximum values of the series resistance $R_\mathrm{s}^{pw}$ and the shunt conductance $G_\mathrm{sh}^{pw}$ are obtained as indicated in \cref{eq:rs_piecewise,eq:gsh_piecewise}. These estimations are based on $\odif[switch-*=true]{V_\mathrm{pv}}{I_\mathrm{pv}}$ for the SDM-3. 
        \begin{equation}
            \label{eq:piecewise_current_A0}
            I_\mathrm{pv} =
            \begin{cases}
                -\cfrac{I_\mathrm{sc}-I_\mathrm{mp}}{V_\mathrm{mp}}V_\mathrm{pv}+I_\mathrm{sc}, & \text{if} \quad V_\mathrm{pv} \in \,  \bigr[  0, \, V_\mathrm{mp} \bigl], \\
                \cfrac{I_\mathrm{mp}}{V_\mathrm{oc}-V_\mathrm{mp}} (V_\mathrm{oc}-V_\mathrm{pv}) , & \text{if} \quad V_\mathrm{pv} \in \,  \bigl]  V_\mathrm{mp}, \, V_\mathrm{oc} \bigr].
            \end{cases}
        \end{equation}

        \begin{align}
            \label{eq:rs_piecewise}
            R_\mathrm{s}^{pw} &= \cfrac{V_\mathrm{oc}-V_\mathrm{mp}}{I_\mathrm{mp}} \\
            \label{eq:gsh_piecewise}
            G_\mathrm{sh}^{pw} &= \cfrac{(I_\mathrm{sc}-I_\mathrm{mp})I_\mathrm{mp}}{V_\mathrm{oc}I_\mathrm{mp}+V_\mathrm{mp}I_\mathrm{sc}-V_\mathrm{oc}I_\mathrm{sc}}
        \end{align}

    \end{remark}

\section{Conclusion} \label{sec:conclusion}
This paper presents a method for computing the SDM boundaries given the cardinal points. From the theoretical part, it is shown that the SDM follows the inequalities $\nicefrac{V_\mathrm{oc}}{2}<V_\mathrm{mp}<V_\mathrm{oc}$ and $\nicefrac{I_\mathrm{sc}}{2}<I_\mathrm{mp}<I_\mathrm{sc}$ (\emph{cf.} \cref{theorem:mppSDM}). Then, it is possible to define implicitly the SDM by means of one single parameter (\emph{cf.} \cref{theorem_one-dimensional}). Under the implicit SDM representation, a computational algorithm is developed to determine the limits of its parameters. To demonstrate the applicability of the algorithm, a Python implementation has been developed and tested using three approaches. 
Firstly, employing a scaled representation of the single-diode model, we generate and utilize $40\,000$ maximum power points to determine $a_\mathrm{max}$.
It's notable that $a_\mathrm{max}$ demonstrates a tendency to increase as the MPP approaches the NRL limit.
Secondly, we analyses the ideality factor of diodes using various photovoltaic technologies including mono-c-Si, multi-c-Si, Thin Film, CdTe, and CIGS.
The results in this section indicate that the maximum diode ideality factor for silicon technology lies predominantly in the range of 1 to 2, with most cases having physical significance.
However, in the case of thin film technologies, CdTe and CIGS, it is difficult to draw conclusive observations due to the limitations imposed by the size of the database.
In the last section of this work, we reported the value of the error metric S-MSD of $38\,929$ I-V curves from the NREL database.
The mean S-MSD value is on the order of $1.34 \times 10^{-4}$ with a standard deviation of $3.03\times 10^{-3}$. 
These consistently low values highlight the practical effectiveness of the SDM-1 approach.


Future work should extend the study of mathematical limitations on other models, \emph{e.g.}, the double-diode model or granular models (two SDM connected in series or connected in parallel). In addition, future studies should investigate the impact of different error functions when determining the optimal parameters combination, \emph{i.e.}, the error on the voltage, the error on the current/voltage, \emph{etc}.
In this context, extending the current methodology to incorporate several cell irradiances and temperatures within the solar PV system is desirable. 
This could lead to the development of a comprehensive general methodology for describing the electrical operation of solar PV systems.





\appendix

\section{Mathematical definitions} 

\subsection{SDM on solar cells} 
\label{app_sdm_solarcell}

If the number of series solar cell $N_\mathrm{s}$ and the parallel solar systems $N_\mathrm{p}$ are known, then it is possible to estimate the SDM of a single solar cell as
\begin{multline}
    \label{eq:solar_cell_estimation}
    \left( V_\mathrm{pv}^\mathrm{cell},\,I_\mathrm{pv}^\mathrm{cell},\,I_\mathrm{ph}^\mathrm{cell},\,I_\mathrm{o}^\mathrm{cell},A^\mathrm{cell},n^\mathrm{cell},R_\mathrm{s}^\mathrm{cell},G_\mathrm{sh}^\mathrm{cell} \right) \coloneqq \\ \left( \frac{V_\mathrm{pv}}{N_\mathrm{s}},\frac{I_\mathrm{pv}}{N_\mathrm{p}},\frac{I_\mathrm{ph}}{N_\mathrm{p}},\frac{I_\mathrm{o}}{N_\mathrm{p}},\frac{A}{N_\mathrm{s}} ,\frac{n}{N_\mathrm{s}},\frac{R_\mathrm{s}  N_\mathrm{p}}{N_\mathrm{s}},\frac{G_\mathrm{sh}  N_\mathrm{s}}{N_\mathrm{p}} \right). 
\end{multline}
For this, the following assumptions are made on the solar PV system: (i) the uniform irradiance, (ii) uniform cell temperature and (iii) uniform state of health.

\subsection{Scaled representation} 
\label{app:scaled}

The parameters of the SDM can be scaled by means of $I_\mathrm{sc}$ and $V_\mathrm{oc}$ \cite{rauschenbach_solar_1980} as
\begin{multline}
    \label{eq:scaled_sdm004}
    (v_{\mathrm{pv}},\,i_{\mathrm{pv}},\,i_{\mathrm{ph}},\,i_{\mathrm{o}},\,a,\,r_{\mathrm{s}},\,g_{\mathrm{sh}}) \coloneqq \\ \left( \frac{V_\mathrm{pv}}{V_\mathrm{oc}},\,\frac{I_\mathrm{pv}}{I_\mathrm{sc}},\,\frac{I_\mathrm{ph}}{I_\mathrm{sc}},\,\frac{I_\mathrm{o}}{I_\mathrm{sc}},\,\frac{A}{V_\mathrm{oc}},   \frac{R_\mathrm{s}  I_\mathrm{sc}}{V_\mathrm{oc}} ,\,  \frac{G_\mathrm{sh}  V_\mathrm{oc}}{I_\mathrm{sc}} \right). 
\end{multline}

\section{Mathematical calculations}

\subsection{Concavity of the SDM} 
\label{app:concavity}

The first and second derivatives of the SDM according to $V_\mathrm{pv}$, are computed as
\begin{align}
    \label{eq:1stderivativeIpv}
    \odif[switch-*=true]{V_\mathrm{pv}}{I_\mathrm{pv}}  & \coloneqq - \frac{I_{\mathrm{o}}  \exp{ \left( \cfrac{V_\mathrm{pv}+R_\mathrm{s}  I_\mathrm{pv}}{A} \right) }+  A  G_{\mathrm{sh}}}{A + R_{\mathrm{s}}  \left( I_{\mathrm{o}}  \exp{ \left( \cfrac{V_\mathrm{pv}+R_\mathrm{s}  I_\mathrm{pv}}{A} \right) }+  A  G_{\mathrm{sh}} \right) }, \\
    \label{eq:2ndderivativeIpv}
    \odif[switch-*=true,order=2]{V_\mathrm{pv}}{I_\mathrm{pv}} & \coloneqq - \frac{ I_{\mathrm{o}}  A  \exp{ \left( \cfrac{V_\mathrm{pv}+R_\mathrm{s}  I_\mathrm{pv}}{A} \right) }}{ \left( A + R_{\mathrm{s}}  \left( I_{\mathrm{o}}  \exp{ \left( \cfrac{V_\mathrm{pv}+R_\mathrm{s}  I_\mathrm{pv}}{A} \right) }+  A  G_{\mathrm{sh}} \right) \right)^3 }. 
\end{align}
Since all five parameters must be positives, it is accomplished that $\odif[switch-*=true]{V_\mathrm{pv}}{I_\mathrm{pv}}<0$ and $\odif[switch-*=true,order=2]{V_\mathrm{pv}}{I_\mathrm{pv}}<0$. Therefore, the SDM is concave down (decreasing) $\forall \, V_\mathrm{pv} \in \mathbb{R}$.

\subsection{Maximum power limits of a simplified I-V curve}
\label{appendix:cp-limits}

Let be a simplified I-V curve formed by the union of two segments at the point $(V_\mathrm{x}, I_\mathrm{x})$ such that $0<\nicefrac{V_\mathrm{x}}{V_\mathrm{oc}}<1$ and $(1-\nicefrac{V_\mathrm{x}}{V_\mathrm{oc}})<\nicefrac{I_\mathrm{x}}{I_\mathrm{sc}}<1$, described as  
\begin{align}
    \label{eq:iv_simplified}
    I_\mathrm{pv}(V_\mathrm{pv})
    &\coloneqq 
    \begin{cases}
        I_\mathrm{sc} - \cfrac{ I_\mathrm{sc}-I_\mathrm{x} }{V_\mathrm{x}} V_\mathrm{pv} , & \text{if} \quad V_\mathrm{pv} \in \,  \bigr[  0, \, V_\mathrm{x} \bigl], \\
        \cfrac{I_\mathrm{x} \left(   V_\mathrm{oc} - V_\mathrm{pv} \right)}{V_\mathrm{oc}-V_\mathrm{x}} , & \text{if} \quad V_\mathrm{pv} \in \,  \bigr]  V_\mathrm{x}, \, V_\mathrm{oc} \bigl].
    \end{cases} 
\end{align}  
Then, the maximum power voltage $V_\mathrm{mp}^\mathrm{x}$ is computed from the photovoltaic power:
\begin{align}
    \label{eq:pv_simplified}
    P_\mathrm{pv}(V_\mathrm{pv})
    &\coloneqq 
    \begin{cases}
        I_\mathrm{sc} V_\mathrm{pv} - \cfrac{ I_\mathrm{sc}-I_\mathrm{x} }{V_\mathrm{x}}  V_\mathrm{pv}^2 , & \text{if} \quad V_\mathrm{pv} \in \,  \bigr[  0, \, V_\mathrm{x} \bigl], \\
        \cfrac{I_\mathrm{x} \left(   V_\mathrm{oc} - V_\mathrm{pv} \right)}{V_\mathrm{oc}-V_\mathrm{x}} , & \text{if} \quad V_\mathrm{pv} \in \,  \bigr]  V_\mathrm{x}, \, V_\mathrm{oc} \bigl],
    \end{cases}
\end{align}
by taking its derivative according to $V_\mathrm{pv}$: 
\begin{align}
    \label{eq:vmpx_simplified}
    V_\mathrm{mp}^\mathrm{x}
    & = 
    \begin{cases}
        \cfrac{V_\mathrm{oc}}{2}, & \text{if} \quad \cfrac{V_\mathrm{x}}{V_\mathrm{oc}} < \cfrac{I_\mathrm{sc}-I_\mathrm{x}}{I_\mathrm{sc}}, \\
        \cfrac{I_\mathrm{sc} V_\mathrm{x}}{2(I_\mathrm{sc}-I_\mathrm{x})} , & \text{if} \quad \cfrac{I_\mathrm{sc}-I_\mathrm{x}}{I_\mathrm{sc}} \leqslant \cfrac{V_\mathrm{x}}{V_\mathrm{oc}} \leqslant   \cfrac{2(I_\mathrm{sc}-I_\mathrm{x})}{I_\mathrm{sc}}, \\
        V_\mathrm{oc}, & \text{if} \quad \cfrac{2(I_\mathrm{sc}-I_\mathrm{x})}{I_\mathrm{sc}} < \cfrac{V_\mathrm{x}}{V_\mathrm{oc}}.
    \end{cases}
\end{align}  
Here, it is noted that $\nicefrac{V_\mathrm{oc}}{2}<V_\mathrm{mp}^\mathrm{x}<V_\mathrm{oc}$. Similarly, the maximum power current $I_\mathrm{x}$ must accomplish $\nicefrac{I_\mathrm{sc}}{2}<I_\mathrm{mp}^\mathrm{x}<I_\mathrm{sc}$.

\begin{figure}[!hbt]
    \centering
    \includegraphics[width=0.75\textwidth]{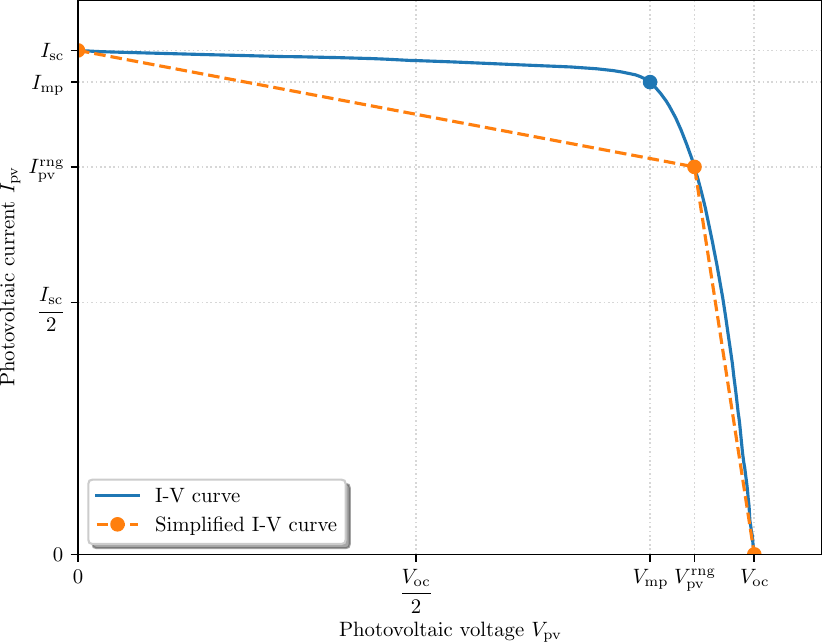}
    \caption{Simplified I-V curve determined from the union of two segments: (i) one connecting the points $(0,\,\,I_\mathrm{sc})$ and $(V_\mathrm{pv}^\mathrm{rng},\,\,I_\mathrm{pv}^\mathrm{rng})$ and (ii) $(V_\mathrm{pv}^\mathrm{rng},\,\,I_\mathrm{pv}^\mathrm{rng})$ and $(V_\mathrm{oc},\,0)$.} \label{fig:iv_curve_rng}
\end{figure}

Now, let's consider a random point $(V_\mathrm{pv}^\mathrm{rng}, I_\mathrm{pv}^\mathrm{rng})$ on an I-V curve. Graphically, \cref{fig:iv_curve_rng} depicts this situation. According to the explained behavior for the simplified I-V curve, the maximum power point of the simplified I-V curve follows $\nicefrac{V_\mathrm{oc}}{2}<V_\mathrm{mp}^\mathrm{rng}<V_\mathrm{mp}$ and $\nicefrac{I_\mathrm{sc}}{2}<I_\mathrm{mp}^\mathrm{rng}<I_\mathrm{mp}$.

\subsection{Maximum power of the simplified NSHM} 
\label{appendix:mp_NSHM}

The simplified current as a function of the voltage is computed from \cref{def:NRLM-1} when $A \rightarrow 0^+$.
Mathematically, it is expressed as
\begin{align}
    \label{eq:iv_simplified_nshm}
    I_\mathrm{pv}(V_\mathrm{pv})
    & \coloneqq 
    \begin{cases}
        I_\mathrm{sc}, & \text{if} \quad V_\mathrm{pv} \in \,  \bigr[  0, \, V_\mathrm{i} \bigl], \\
        \cfrac{V_\mathrm{oc}-V_\mathrm{pv}}{R_\mathrm{s}} , & \text{if} \quad V_\mathrm{pv} \in \,  \bigr]  V_\mathrm{i}, \, V_\mathrm{oc} \bigl].
    \end{cases}
\end{align}  
This representation of the NSHM is continuous at the inflection point $V_\mathrm{i}=V_\mathrm{oc}-R_\mathrm{s}I_\mathrm{sc}$. 
Then, the photovoltaic power obtained from this curve must be a continuous function, which is composed of a linear component and a quadratic component: 
\begin{align}
    \label{eq:pv_simplified_nshm}
    P_\mathrm{pv}(V_\mathrm{pv})
    &\coloneqq \begin{cases}
    I_\mathrm{sc}  V_\mathrm{pv}, & \text{if} \quad V_\mathrm{pv} \in \,  \bigr[  0, \, V_\mathrm{i} \bigl], \\
    \cfrac{(V_\mathrm{oc}-V_\mathrm{pv})  V_\mathrm{pv} }{R_\mathrm{s}} , & \text{if} \quad V_\mathrm{pv} \in \,  \bigr]  V_\mathrm{i}, \, V_\mathrm{oc} \bigl].
    \end{cases}
\end{align}
From this, the maximum power voltage must correspond either to the intersection between the linear and quadratic sections or to the maximum value of the quadratic section. In this sense, the maximum power voltage as a function of the series resistance is indicated in \cref{eq:vmp_simplified_nshm}. 

\begin{align}
V_\mathrm{pv}^\mathrm{NSH}(R_\mathrm{s})
&\coloneqq \begin{cases}
V_\mathrm{oc}-R_\mathrm{s} I_\mathrm{sc}, & \text{if} \quad R_\mathrm{s} \in \,  \biggr]  0, \, \cfrac{V_\mathrm{oc}}{2 I_\mathrm{sc}} \biggl[, \\
\cfrac{V_\mathrm{oc}}{2 I_\mathrm{sc}} , & \text{if} \quad R_\mathrm{s} \in \,  \biggr[  \cfrac{V_\mathrm{oc}}{2}, \, \cfrac{V_\mathrm{oc}}{I_\mathrm{sc}} \biggl[.
\end{cases} \label{eq:vmp_simplified_nshm}
\end{align} 

Then evaluating the maximum power voltage in \cref{eq:iv_simplified_nshm}, the maximum power current can be found. 

\subsection{Maximum power of the simplified NSRM} 
\label{appendix:mp_NSRM}

The simplified current as a function of the voltage is presented in \cref{eq:iv_simplified_nsrm}. 
\begin{align}
\label{eq:iv_simplified_nsrm}
I_\mathrm{pv}(V_\mathrm{pv})
&\coloneqq \begin{cases}
I_\mathrm{sc} - G_\mathrm{sh}  V_\mathrm{pv}, & \text{if} \quad  V_\mathrm{pv} \in \,  \bigr[  0, \, V_\mathrm{oc} \bigl[, \\
0 , & \text{if} \quad V_\mathrm{pv} = V_\mathrm{oc}.
\end{cases}
\end{align} 
This curve is obtained for the case where $A \rightarrow 0^+$. This representation is continuous for $V_\mathrm{pv} \in \left[ 0, \, V_\mathrm{oc} \right[$. Then, the photovoltaic power is represented as indicated by \cref{eq:pv_simplified_nsrm}. 
\begin{align}
\label{eq:pv_simplified_nsrm}
P_\mathrm{pv}(V_\mathrm{pv})
&\coloneqq \begin{cases}
(I_\mathrm{sc} - G_\mathrm{sh}  V_\mathrm{pv})  V_\mathrm{pv} , & \text{if} \quad  V_\mathrm{pv} \in \,  \bigr[  0, \, V_\mathrm{oc} \bigl[, \\
0 , & \text{if} \quad V_\mathrm{pv} = V_\mathrm{oc}.
\end{cases}
\end{align} 
From this expression, the maximum power voltage corresponds to the maximum value of the quadratic function such that $V_\mathrm{mp}^\mathrm{NSH}<V_\mathrm{oc}$. Then, the maximum power voltage is described as indicated by \cref{eq:vmp_simplified_nsrm}. 
\begin{align}
\label{eq:vmp_simplified_nsrm}
V_\mathrm{mp}^\mathrm{NSR}(G_\mathrm{sh})
&\coloneqq \begin{cases}
V_\mathrm{oc} , & \text{if} \quad  G_\mathrm{sh} \in \,  \biggr]  0, \, \cfrac{I_\mathrm{sc}}{2V_\mathrm{oc}} \biggl[, \\
\cfrac{I_\mathrm{sc}}{2 G_\mathrm{sh}}, & \text{if} \quad G_\mathrm{sh}  \in \,  \biggr[  \cfrac{I_\mathrm{sc}}{2V_\mathrm{oc}}, \, \cfrac{I_\mathrm{sc}}{V_\mathrm{oc}} \biggl[.
\end{cases}
\end{align} 
Regarding the maximum power current, this value can be obtained from the direct evaluation of $V_\mathrm{mp}^\mathrm{NSH}$ on \cref{eq:iv_simplified_nsrm}.

\subsection{Three-parameters SDM} \label{appendix:3D_SDM}
To determine $F_\mathrm{SD3}(X;\,\Theta_3)$, explicit expressions of $I_\mathrm{ph}(\Theta_3)$ and $I_\mathrm{o}(\Theta_3)$ are required. For this, the short circuit current and the open circuit voltage are evaluated in \cref{eq:sdm_a}. Then, it is possible to write the equation system shown in \cref{eq:system3DSDMa,eq:system3DSDMb}. The solution of this equation system is indicated in \cref{eq:system02a,eq:system02b,eq:system02c}. Replacing these parameters in \cref{def:SDM}, the expression of $F_\mathrm{SD3}(X;\,\Theta_3)$ is obtained and presented in \cref{eq:sdmmodel003}. 
Since $I_\mathrm{ph}(\Theta_3)$ and $I_\mathrm{o}(\Theta_3)$ must be positives, some novel restrictions on $R_\mathrm{s}$ and $G_\mathrm{sh}$ are included. Firstly, from $I_\mathrm{o}(\Theta_3)$ it is noted that $R_\mathrm{s}<\nicefrac{V_\mathrm{oc}}{I_\mathrm{sc}}$. Otherwise, if $R_\mathrm{s}>\nicefrac{V_\mathrm{oc}}{I_\mathrm{sc}}$ implies $I_\mathrm{sc} -G_\mathrm{sh}  ( V_\mathrm{oc} - R_\mathrm{s}  I_\mathrm{sc}) > 0$ and therefore $I_\mathrm{o}(\Theta_3)<0$. Then, it must accomplished be $G_\mathrm{sh}<\nicefrac{I_\mathrm{sc}}{V_\mathrm{oc}-R_\mathrm{s}  I_\mathrm{sc}}$. No further information can be obtained from $I_\mathrm{ph}(\Theta_3)$.
%
\begin{align}
\label{eq:system3DSDMa}
I_\mathrm{ph} + \left(  1 - \exp{\left( \cfrac{R_\mathrm{s}  I_\mathrm{sc}}{A} \right)} \right) I_\mathrm{o} &=  \left(  R_\mathrm{s} G_\mathrm{sh} + 1 \right) I_\mathrm{sc}  \\
\label{eq:system3DSDMb}
I_\mathrm{ph} + \left( 1 - \exp{\left( \cfrac{V_\mathrm{oc}}{A} \right)} \right) I_\mathrm{o} &= G_\mathrm{sh} V_\mathrm{oc}
\end{align}
\begin{align}
\label{eq:system02a}
I_\mathrm{ph}\left( \Theta_3 \right) &\coloneqq \cfrac{F_\mathrm{ph}(\Theta_3) }{\exp{ \left(\cfrac{V_\mathrm{oc}}{A} \right)}-\exp{\left(\cfrac{R_\mathrm{s}  I_\mathrm{sc}}{A}\right)}} \\
\label{eq:system02b}
I_{\mathrm{o}}( \Theta_3 ) &\coloneqq \frac{ I_\mathrm{sc}-G_{\mathrm{sh}}  \left(V_\mathrm{oc}- R_{\mathrm{s}}  I_\mathrm{sc} \right) }{\exp{ \left( \cfrac{V_\mathrm{oc}}{A} \right)}-\exp{\left(\cfrac{R_\mathrm{s}  I_\mathrm{sc}}{A}\right)}} 
\end{align}
\begin{multline}
\label{eq:system02c}
F_\mathrm{ph}(\Theta_3) \coloneqq  ( R_\mathrm{s}G_\mathrm{sh}+1) I_\mathrm{sc}   \exp{ \left(\cfrac{V_\mathrm{oc}}{A} \right)} \\ -G_\mathrm{sh}  V_\mathrm{oc}  \exp{\left(\cfrac{R_\mathrm{s}  I_\mathrm{sc}}{A}\right)}  - ( R_\mathrm{s}G_\mathrm{sh}+1) + G_\mathrm{sh} V_\mathrm{oc}
\end{multline}

\begin{multline}
F_\mathrm{SD3}(X;\,\Theta_3) \coloneqq (-I_\mathrm{pv} (  R_\mathrm{s}G_\mathrm{sh}+1 ) + G_\mathrm{sh}  (V_\mathrm{oc}-V_\mathrm{pv}) )  \exp{\left( \cfrac{R_\mathrm{s} I_\mathrm{sc}}{A} \right)} \\ +(I_\mathrm{sc}(R_\mathrm{s}G_\mathrm{sh}+1)-G_\mathrm{sh}V_\mathrm{oc})  \exp{\left( \cfrac{V_\mathrm{pv}+R_\mathrm{s} I_\mathrm{pv}}{A} \right)} \\ + ((I_\mathrm{pv}-I_{\mathrm{sc}})  (R_\mathrm{s} G_\mathrm{sh}+1 )+G_{\mathrm{sh}}  V_{\mathrm{pv}})  \exp{\left( \cfrac{V_\mathrm{oc}}{A} \right)}
\label{eq:sdmmodel003}
\end{multline}


\subsection{Null resistive losses limit} \label{app:nrm_limit}
The determination of the NRL limit uses the definition of the NRML-1 given in \cref{def:NRLM-1}. Then, to find the NRL limit, the equation system presented in \cref{eq:nrlm_eqsyst01,eq:nrlm_eqsyst02} must be solved. Here, \cref{eq:nrlm_eqsyst01} represents the NRLM evaluated at $(V_\mathrm{pv},\,I_\mathrm{pv})=(V_\mathrm{mp}^\mathrm{NRL},\,I_\mathrm{mp}^\mathrm{NRL})$. On the other hand, \cref{eq:nrlm_eqsyst02} represents the derivative of the current according to the voltage evaluated at the MPP. The evaluation of this derivative function is made according to \cref{lemma:photovoltaic_mpp}. 
\begin{align}
    0 &= -(I_\mathrm{sc}-I_\mathrm{mp}^\mathrm{NRL})  \exp{\left( \frac{V_\mathrm{oc}}{A} \right)} + I_\mathrm{sc}  \exp{\left( \frac{V_\mathrm{mp}^\mathrm{NRL}}{A} \right)} - I_\mathrm{mp}^\mathrm{NRL} \label{eq:nrlm_eqsyst01}  \\ 
    0 &= A  I_\mathrm{mp}^\mathrm{NRL}  \left( \exp{ \left( \cfrac{V_\mathrm{oc}}{A}-1 \right)} \right) - I_\mathrm{sc}  V_\mathrm{mp}^\mathrm{NRL}  \exp{ \left( \cfrac{V_\mathrm{mp}^\mathrm{NRL}}{A} \right)}  \label{eq:nrlm_eqsyst02}
\end{align} 
The equation system accepts different solutions depending on the independent variable selected. In this sense, if $A$ is selected as the independent variable, the solution is indicated in \cref{eq:NRLM_voltage_maximum,eq:NRLM_current_maximum}. Another solution is to select $I_\mathrm{mp}^\mathrm{NRL}$ as the independent variable, as indicated by \cref{eq:vNRLM_impp_c,eq:vNRLM_impp_d}.
\begin{equation}
    z_\mathrm{mp} = \cfrac{I_\mathrm{mp}^\mathrm{NRL} }{I_\mathrm{mp}^\mathrm{NRL}-I_\mathrm{sc}}  \exp{ \left( \cfrac{I_\mathrm{mp}^\mathrm{NRL}}{I_\mathrm{mp}^\mathrm{NRL}-I_\mathrm{sc}} \right) } \label{eq:vNRLM_impp_d}
\end{equation}
\begin{multline}
    V_\mathrm{mp}^\mathrm{NRL}(I_\mathrm{mp}^\mathrm{NRL}) 
    = \\ \cfrac{ V_\mathrm{oc}  \left( W_0 \left(   \cfrac{ e  I_\mathrm{mp}^\mathrm{NRL}  \left( \left( I_\mathrm{mp}^\mathrm{NRL} - I_\mathrm{sc} \right)  W_0 \left( z_\mathrm{mp} \right) - I_\mathrm{sc} \right) }{ \left( I_\mathrm{mp}^\mathrm{NRL}-I_\mathrm{sc} \right)^2  W_0 \left( z_\mathrm{mp} \right) }  \right)  - 1 \right) }{ \ln{ \left( \cfrac{ I_\mathrm{mp}^\mathrm{NRL}  \left( \left( I_\mathrm{mp}^\mathrm{NRL} - I_\mathrm{sc} \right)  W_0 \left(  z_\mathrm{mp} \right) - I_\mathrm{sc} \right) }{ \left( I_\mathrm{mp}^\mathrm{NRL}-I_\mathrm{sc} \right)^2  W_0 \left(  z_\mathrm{mp} \right) } \right)}} \label{eq:vNRLM_impp_c}
\end{multline}
\subsection{Two-parameter SDM} \label{app:two-parameter_SDM}
\begin{figure}[!hbt]
\centering
\includegraphics[width=0.75\textwidth]{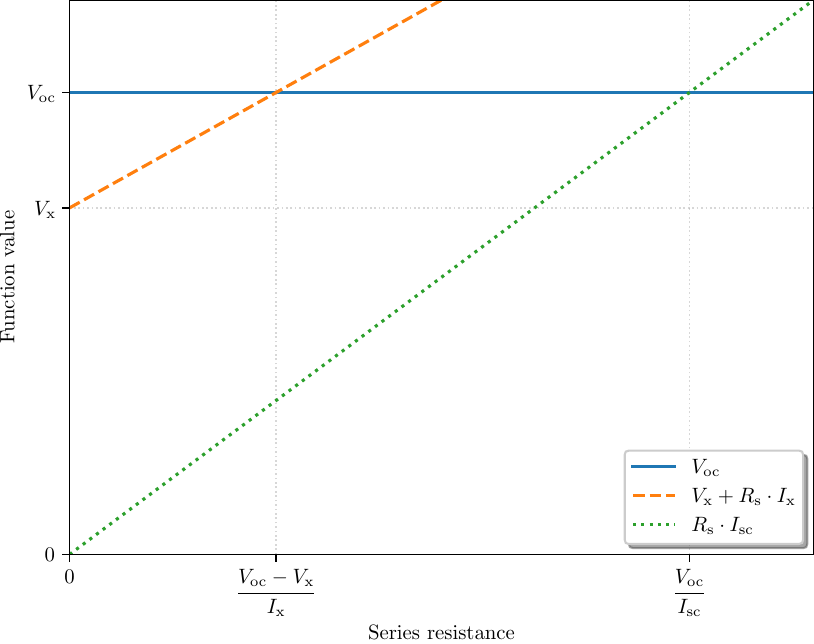}
\caption{Restrictions on $R_\mathrm{s}$ for the two-parameter SDM. The blue line indicates the open circuit voltage $V_\mathrm{oc}$, the dashed orange line indicates the voltage drop at the point $(V_\mathrm{x},\,I_\mathrm{x})$ and the dotted green line indicates the voltage drop at the short circuit current.} \label{fig:rsmax_limits}
\end{figure}

To determine $F_\mathrm{SD2}(X;\,\Theta_2)$, explicit expressions of $I_\mathrm{ph}(\Theta_2)$, $I_\mathrm{o}(\Theta_2)$ and $G_\mathrm{sh}(\Theta_2)$ are required. For this, the short circuit current, the open circuit voltage and a point $(V_\mathrm{x},\,I_\mathrm{x})$ selected according to \cref{lemma:concavity} are evaluated in \cref{eq:sdm_a}. Then, it is possible to write the equation system shown in \cref{eq:system2DSDMa,eq:system2DSDMb,eq:system2DSDMc}. The solution of this equation system is indicated in \cref{eq:2D_sdm001,eq:2D_sdm002,eq:2D_sdm003,eq:2D_sdm006_ph,eq:2D_sdm004,eq:2D_sdm005}. Replacing these parameters in \cref{def:SDM}, the expression of $F_\mathrm{SD2}(X;\,\Theta_2)$ is obtained and presented in \cref{eq:2D_sdm_Fsd2}.
\begin{equation}
    \label{eq:system2DSDMa}
    I_\mathrm{ph} + \left( 1- \exp{\left( \cfrac{R_\mathrm{s}  I_\mathrm{sc}}{A} \right)} \right) I_\mathrm{o} - R_\mathrm{s} I_\mathrm{sc} G_\mathrm{sh} = I_\mathrm{sc}
\end{equation}
\begin{equation}
    \label{eq:system2DSDMb}
    I_\mathrm{ph} + \left( 1- \exp{\left( \cfrac{V_\mathrm{oc}}{A} \right)} \right) I_\mathrm{o} - V_\mathrm{oc} G_\mathrm{sh} = 0 
\end{equation}
\begin{equation}
    \label{eq:system2DSDMc}
    I_\mathrm{ph} + \left( 1- \exp{\left( \cfrac{V_\mathrm{x}+R_\mathrm{s} I_\mathrm{x}}{A} \right)} \right) I_\mathrm{o} - (V_\mathrm{x}+R_\mathrm{s} I_\mathrm{x}) G_\mathrm{sh} = I_\mathrm{x}
\end{equation}
\begin{align}
    I_\mathrm{ph}(\Theta_2) &\coloneqq \cfrac{ F_\mathrm{ph}(\Theta_2) }{F_\mathrm{cd}(\Theta_2)} \label{eq:2D_sdm001} \\
    I_\mathrm{o}(\Theta_2) &\coloneqq \frac{V_\mathrm{x} I_\mathrm{sc}+V_\mathrm{oc} I_\mathrm{x}-V_\mathrm{oc} I_\mathrm{sc}}{F_\mathrm{cd}(\Theta_2)} \label{eq:2D_sdm002} \\
    G_\mathrm{sh}(\Theta_2) & \coloneqq \frac{F_\mathrm{sh}(\Theta_2)}{F_\mathrm{cd}(\Theta_2)} \label{eq:2D_sdm003}
\end{align}
\begin{multline}
    \label{eq:2D_sdm006_ph}
    F_\mathrm{ph}(\Theta_2) \coloneqq -(V_\mathrm{x}  I_\mathrm{sc}+V_\mathrm{oc} I_\mathrm{x} - V_\mathrm{oc}  I_\mathrm{sc}) + V_\mathrm{x} I_\mathrm{sc} \exp{\left( \frac{V_\mathrm{oc}}{A} \right)} \\ - V_\mathrm{oc} I_\mathrm{sc}  \exp{\left( \frac{V_\mathrm{x}+R_\mathrm{s}  I_\mathrm{x}}{A} \right)} + V_\mathrm{oc} I_\mathrm{x}  \exp{\left( \frac{R_\mathrm{s}  I_\mathrm{sc}}{A} \right)}
\end{multline}
\begin{multline}
    \label{eq:2D_sdm004}
    F_\mathrm{sh}(\Theta_2) \coloneqq (I_\mathrm{sc}-I_\mathrm{x})  \exp{\left( \frac{V_\mathrm{oc}}{A} \right)} \\ - I_\mathrm{sc}  \exp{\left( \frac{V_\mathrm{x}+R_\mathrm{s}  I_\mathrm{x}}{A} \right)} + I_\mathrm{x}  \exp{\left( \frac{R_\mathrm{s}  I_\mathrm{sc}}{A} \right)}
\end{multline}
\begin{multline}
    \label{eq:2D_sdm005}
    F_\mathrm{cd}(\Theta_2) \coloneqq (V_\mathrm{x}-R_\mathrm{s}(I_\mathrm{sc}-I_\mathrm{x}) ) \exp{\left( \frac{V_\mathrm{oc}}{A} \right)} \\ - (V_\mathrm{oc} -R_\mathrm{s} I_\mathrm{sc}) \exp{\left( \frac{V_\mathrm{x}+R_\mathrm{s}  I_\mathrm{x}}{A} \right)} \\ + (V_\mathrm{oc}-V_\mathrm{x} - R_\mathrm{s} I_\mathrm{x})  \exp{\left( \frac{R_\mathrm{s}  I_\mathrm{sc}}{A} \right)} 
\end{multline}
%
\begin{multline}
    \label{eq:2D_sdm_Fsd2}
    F_\mathrm{SD2}(X;\,\Theta_2) \coloneqq ( V_\mathrm{oc}  I_\mathrm{sc} -V_\mathrm{x}  I_\mathrm{sc} - V_\mathrm{oc}  I_\mathrm{x} )  \exp{\left( \cfrac{V_\mathrm{pv}+R_\mathrm{s} I_\mathrm{pv}}{A} \right)} \\ + (V_\mathrm{x}  ( I_\mathrm{sc} - I_\mathrm{pv} ) - V_\mathrm{pv}  ( I_\mathrm{sc} - I_\mathrm{x} ))  \exp{\left( \cfrac{V_\mathrm{oc}}{A} \right)}  \\ + (V_\mathrm{pv}  I_\mathrm{sc} + V_\mathrm{oc}  I_\mathrm{pv} - V_\mathrm{oc}  I_\mathrm{sc} )  \exp{\left( \cfrac{V_\mathrm{x}+R_\mathrm{s} I_\mathrm{x}}{A} \right)} \\  + (I_\mathrm{x}  ( V_\mathrm{oc} - V_\mathrm{pv} ) - I_\mathrm{pv}  ( V_\mathrm{oc} - V_\mathrm{mp} ))  \exp{\left( \cfrac{R_\mathrm{s} I_\mathrm{sc}}{A} \right)}
\end{multline}
%
%
According to \cref{prop:2D_restriction}, it must be ensured $F_\mathrm{sh}(\Theta_2)>0$. For this, $R_\mathrm{s}  I_\mathrm{sc}<V_\mathrm{x}+R_\mathrm{s}  I_\mathrm{x}< V_\mathrm{oc}$ must be satisfied, otherwise $F_\mathrm{sh}(\Theta_2)$ becomes negative. This situation is depicted graphically in \cref{fig:rsmax_limits}. Here, a novel restriction on $R_\mathrm{s}$ is included as following $R_\mathrm{s}<\nicefrac{(V_\mathrm{oc}-V_\mathrm{x})}{I_\mathrm{x}}$.
\subsection{MPP of the SDM-2} 
\label{app:mpp_sdm-2}
The maximum power point for the SDM-2 satisfying the cardinal points follows \cref{eq:fmp_theta2}.
\begin{multline}
\label{eq:fmp_theta2} 
F_\mathrm{mp}(\Theta_2) \coloneqq -A  V_\mathrm{mp}  ( 2  I_\mathrm{mp} - I_\mathrm{sc})  \exp{\left( \cfrac{V_\mathrm{oc}}{A} \right)} \\ + \left( ( V_\mathrm{oc}  I_\mathrm{mp} + V_\mathrm{mp}  I_\mathrm{sc}  - V_\mathrm{oc}  I_\mathrm{sc} )  (V_\mathrm{mp} - R_\mathrm{s}  I_\mathrm{mp}) \right. \\ \left. +  A \, ( V_\mathrm{oc}  I_\mathrm{mp} - V_\mathrm{mp}  I_\mathrm{sc} ) \right)  \exp{\left( \cfrac{V_\mathrm{mp}+R_\mathrm{s} I_\mathrm{mp}}{A} \right)} \\ + A \, I_\mathrm{mp} \, ( 2  V_\mathrm{mp} - V_\mathrm{oc}) \, \exp{\left( \cfrac{R_\mathrm{s} I_\mathrm{sc}}{A} \right)}
\end{multline}
%
%
\section{NREL database} \label{app:nrel}
The statistical information of the histogram indicated in \cref{fig:nrel_histogram} is addressed by \cref{tab:nrel_histogram_stats}. 
%
\begin{table}[!ht]
\centering
\caption{Detailed statistical information for the I-V curves from the module mSi460A8 installed at Cocoa, Florida in the 2011-2012 period. All quantities mentioned in this table are dimensionless.  Q1, Q2 and Q3 indicate quartiles 1, 2 and 3.  }
\label{tab:nrel_histogram_stats}
\begin{tabular}{ll} \toprule
Number of data & $38\,929$ \\
Mean           & $-1.34\times10^{-4}$ \\
Standard deviation & $3.03\times10^{-3}$ \\
Minimum & $-5\times 10^{-2}$ \\
Q1 & $-1.68\times 10^{-3}$ \\
Q2 & $-1.15\times 10^{-3}$ \\
Q3 & $2.07 \times 10^{-4}$ \\
Maximum & $1.10 \times 10^{-2}$ \\ \bottomrule
\end{tabular}
\end{table}

\bibliographystyle{elsarticle-num-names} 

\bibliography{references}





\end{document}